\documentclass[onecolumn,twoside]{IEEEtran}
\usepackage[T1]{fontenc}
\usepackage{authblk}
\usepackage{cite}
\usepackage{amsmath,amssymb,amsfonts,dsfont}
\usepackage{algorithm}
\usepackage[noend]{algorithmic}
\usepackage{graphicx}
\usepackage{textcomp}
\usepackage{xcolor}
\usepackage{bm}
\usepackage{booktabs}
\usepackage{multirow}
\usepackage{lipsum}
\usepackage{url}
\usepackage{dashrule}
\usepackage{lipsum}
\newtheorem{theorems}{\textbf{Theorem}}
\newtheorem{proposition}{\textbf{Proposition}}
\newtheorem{remark}{\textbf{Remark}}
\newtheorem{lemma}{\textbf{Lemma}}
\newtheorem{assp}{\textbf{Assumption}}
\newenvironment{proofs}{{\indent \it \textbf{Proof}:}}{\hfill $\blacksquare$\par}

\DeclareMathOperator*{\argmin}{\arg\min} 
\newcommand{\thatis}{\textit{i.e.}}

\newcommand{\mbbR}{\mathbb{R}} 

\newcommand{\bdphi}{\boldsymbol{\phi}} 
\newcommand{\bdpsi}{\boldsymbol{\psi}}

\newcommand{\bdx}{\boldsymbol{x}} 
\newcommand{\bdy}{\boldsymbol{y}}

\newcommand{\bdD}{\boldsymbol{D}}

\newcommand{\bda}{\boldsymbol{\alpha}} 
\newcommand{\bdb}{\boldsymbol{\beta}}

\newcommand{\mcX}{\mathcal{X}}

\newcommand{\mcY}{\mathcal{Y}}

\newcommand{\mrmd}{\mathrm{d}}
\newcommand{\mrme}{\mathrm{e}}

\makeatletter
\renewcommand{\maketag@@@}[1]{\hbox{\m@th\normalsize\normalfont#1}}
\makeatother

\ifCLASSINFOpdf
\else
\fi

\begin{document}
\title{
An Optimal Transport-Based Method for Computing LM Rate and Its Convergence Analysis
}

\author{
    Shitong~Wu,
    Wenhao~Ye,
    Xinwei~Li,
    Lingyi~Chen,
    Wenyi~Zhang, \IEEEmembership{Senior Member,~IEEE,} ~~~~~~~~~~~~~~~~
    Huihui~Wu, \IEEEmembership{Member,~IEEE,}
    and~Hao~Wu
    \thanks{Preliminary results of this work have been presented in part at the 2022 IEEE Global Communications Conference (GLOBECOM) \cite{ye2022optimal}.}
    \thanks{This work was partially supported by National Natural Science Foundation of China (Grants 12271289 and 62231022).
    \textit{(Corresponding author: Wenyi~Zhang.) }
    }
    \thanks{Shitong Wu, Wenhao Ye, Xinwei Li, Lingyi Chen, and Hao Wu are with the Department of Mathematical Sciences, Tsinghua University, Beijing 100084, China.
    }
    \thanks{Wenyi Zhang is with the Department of Electronic Engineering and Information Science, 
    University of Science and Technology of China, Hefei, Anhui 230027, China (e-mail: wenyizha@ustc.edu.cn).
    }
    \thanks{Huihui Wu is with Ningbo Institute of Digital Twin, Eastern Institute of Technology, Zhejiang 315000, China.
    }
    \thanks{The first three authors contributed equally to this work.
    }
}

\markboth{~}%
{~}

\maketitle

\begin{abstract}
The mismatch capacity characterizes the highest information rate of the channel under a prescribed decoding metric and serves as a critical performance indicator in numerous practical communication scenarios.
Compared to the commonly used Generalized Mutual Information (GMI), the 
Lower bound on the Mismatch capacity (LM rate)
generally provides a tighter lower bound on the mismatch capacity.
However, the efficient computation of the LM rate is significantly more challenging than that of the GMI, particularly as the size of the channel input alphabet increases. 
This growth in complexity renders standard numerical methods (e.g., interior point methods) computationally intensive and, in some cases, impractical.
In this work, we reformulate the computation of the LM rate as a special instance of the optimal transport (OT) problem with an additional constraint. Building on this formulation, we develop a novel numerical algorithm based on the Sinkhorn algorithm, which is well known for its efficiency in solving entropy regularized optimization problems.
We further provide the convergence analysis of the proposed algorithm, revealing that the algorithm has a sub-linear convergence rate.
Numerical experiments demonstrate the feasibility and efficiency of the proposed algorithm for the computation of the LM rate. 
\end{abstract}


%

\IEEEpeerreviewmaketitle

\section{Introduction}
%
%
%
%

The
Shannon capacity of a channel defines the maximum achievable transmission rate for reliable communication \cite{Shannon1948}. 
This concept has formed the theoretical foundation of communication theory, continuously inspiring research innovations and informing system design principles for decades.

In many practical scenarios, perfect channel knowledge may not be available or may not be fully leveraged in transceiver designs. 
Notable examples include channels with uncertainty (e.g., fading in wireless communication systems) \cite{2002Fading}, non-ideal transceiver hardware \cite{zhang2011general}, and constrained receiver structures \cite{salz1995com}. 
A common approach in such cases is to employ a fixed decoding metric at the receiver that may not be matched to the actual channel transition probability.
To account for this, the concept of mismatch capacity has been introduced to characterize the maximum achievable information rate under a prescribed decoding metric (see e.g., \cite{1998Reliable,2020Information} and references therein).
Mismatched decoding has been extensively applied in diverse communication scenarios, including bit-interleaved coded modulation (BICM) \cite{Mar2009BICM}, finite-precision arithmetic \cite{BI1999BI,salz1995com}, optical communication \cite{Ghozlan2017Wiener}, and fading channel analysis \cite{2002Fading,2004MultiAn}.
Beyond the classical discrete memoryless channel setting, mismatched decoding has been studied in broader cases, such as source coding \cite{1997Lapi,2019Zhou}, oblivious relaying \cite{2023relaying}, and estimation of zero undetected error capacity \cite{Ah1996Erasure}.

The exact characterization of the mismatch capacity remains an open problem although several lower bounds and upper bounds
\footnote{The gap between these lower bounds and the true mismatch capacity has yet to be conclusively determined. Additionally, some works have addressed upper bounds on mismatch capacity; see, e.g. \cite{2019upper,2022upper}. 
These developments, however, lie beyond the scope of this paper.} 
have been proposed
\cite{1995Channel}.
Among these, the generalized mutual information (GMI) is a relatively simple and widely adopted lower bound \cite{1993Information}, with broad applicability across diverse communication systems \cite{2002Fading,zhang2011general,2012Nearest}.
In contrast, the LM rate 
\footnote{To the best of our knowledge, the term "LM rate", first introduced in \cite{Merhav1994On}, 
appears to stand for the Lower bound on the Mismatch capacity.} 
\cite{Merhav1994On} 
provides a generally tighter lower bound
\footnote{The LM rate is not the tightest possible lower bound. Various improvements upon the GMI and LM rate have been proposed by employing more structured codebook ensembles; see \cite{2020Information}.}
by replacing the independent and identically distributed (i.i.d.) codebook ensemble for the GMI by constant-composition codebook ensemble.

From the perspective of optimization, the primal formulations of the GMI and the LM rate are structurally similar, with the key distinction being an additional constraint in the LM rate formulation: the marginal distribution of the optimizing joint probability distribution over the input and output alphabets need to match a fixed input distribution; see e.g., \cite[Thm. 1]{2000Mismatched}.
Accordingly, the dual form of the GMI is computationally tractable, as it reduces to maximization over a single real variable \cite[Eqn. (12)]{2000Mismatched}, which can be efficiently solved via a one-dimensional line search.
In contrast, the dual form of the LM rate further involves the maximization over variables of the channel input size \cite[Eqn. (11)]{2000Mismatched}, making the problem significantly more challenging.
Thus, relatively little attention has been devoted to the numerical computation of the LM rate.
While general-purpose convex optimization tools such as interior-point methods (e.g., \cite{sdpt3,2010cvx}) can be employed, such processes often require expensive memory consumptions and substantial computational demands.
This motivates the need for dedicated and efficient numerical algorithms for computing the LM rate.

In this work, we reformulate the LM rate as the 
optimal transport (OT) problem with an additional capacity constraint, an approach initially proposed in our preliminary study \cite{ye2022optimal}.
Inspired by OT theory \cite{sinkhorn1967diagonal, 2013sinkhorn}, we present a novel modification to the Sinkhorn algorithm that enables efficient computation of the LM rate based on its constrained optimal transport formulation.
Specifically, each iteration of the proposed algorithm consists of two steps: first, updating transport variables using the standard Sinkhorn update, and second, 
updating the dual variable of the additional constraint with the gradient projection.
Due to the mutual information in the objective function, no additional regularization is required, which ensures convergence within several hundred iterations, in contrast to tens of thousands of iterations typically required for classical OT problems.
%
Moreover, it should be noted that recent works have increasingly explored the connection between information theory and optimal transport \cite{Hayashi2023, LeiHB23,  Yikun10017280, Yang2023, Dor10681556 }.
%

We further analyze the convergence of the proposed method and establish that the algorithm converges at a sub-linear rate by providing an estimate for the error bound.
%
%
%
The primary difficulty stems from an additional constraint inherent to our problem, demanding analytical treatment beyond the classical OT scope. 
Established routines from related constrained OT settings, such as the capacity-constrained OT \cite{Wu2022TheDR} or the rate-distortion problem \cite{wu2022communication}, may inapplicable due to fundamental differences in the distinct constraint structure in the LM rate. 
To overcome this, we leverage a key insight regarding inherent properties of Sinkhorn iterations \cite{ 2018Dvurechensky, HuangJMLR24}.
This insight enables the reduction of our original three-phase alternating optimization scheme to a two-phase process during the convergence analysis. 
Building upon this methodological principle and adapting it to the specific constraint structure, we extend the applicability of the original framework. 
This ultimately allows us to bound the error by controlling two alternating variables. 

The remaining part of this paper is organized as follows. 
Section \ref{sec:back} presents the problem formulations for the LM rate and the OT problem, while Section \ref{sec:lmot} introduces the OT-based LM rate formulation.
In Section \ref{sec:sinkhorn}, we propose a Sinkhorn-type algorithm for solving the LM rate, with the analysis of its convergence rate in Section \ref{sec:analy}.
Numerical experiments are discussed in Section \ref{sec:numer}.
Finally, Section \ref{sec:conclusion} provides the conclusion of this work.

\section{Preliminary} \label{sec:back}
\subsection{Notations}
For clarity, notations are explained in advance. 
The identity matrix of dimension $M$ is denoted by $\bm I_{M}$,
whereas the $M$-dimensional all-one vector is represented by $\mathbf{1}_{M}$.
All vectors are assumed to be column vectors unless otherwise specified.
Vectors are denoted by bold lower case letters, such as $\bda$, $\bdy$, while their corresponding elements are denoted by lower case letters in normal type with indices, such as $\alpha_{i}$, $y_j$. 
Similarly, matrices are written in bold uppercase letters like $\boldsymbol{\Lambda}$, with their elements indicated by the corresponding normal-sized indices $\Lambda_{ij}$.
Meanwhile, uppercase letters of normal type represent constants, such as $T$. 
The superscript $\ell$ indicates that the variable pertains to the $\ell$-th iteration of the algorithm, as shown in $\bda^{\ell}$. 
Moreover, the notions $\odot$ and $\otimes$ represent the Hadamard product and the Kronecker product, respectively. 
The gradient with respect to $\bda$ is denoted by $\nabla_{\bda}$. 
And for a real number $x$, $[x]_{+} \triangleq \max\{x,0\}$.

\subsection{LM Rate}
We consider a discrete memoryless communication channel (DMC) model with a transition law $W(y|x)$ over the finite and discrete channel input alphabet $\mcX= \{x_1,\cdots,x_M\}\subset\mbbR^2$ and channel output alphabet $\mcY \subset\mbbR^2$.
Given a probability measure $P_{X}$ on $\mcX$, we can define the joint probability distribution $P_{XY}$ on $\mcX\times\mcY$ and the output distribution $P_{Y}$ on $\mcY$ as follows
\begin{align*}
    P_{XY}(x_i,B)  = W(B|x_i)P_{X}(x_i), ~ \forall B\subset\mcY, \\
    P_{Y}(B)  = \sum_{i=1}^{M} W(B|x_{i})P_{X}(x_i), ~ \forall B\subset\mcY.
\end{align*}
This discrete memoryless channel is characterized by a probabilistic mapping from the input alphabet $\mcX$ to output alphabet $\mcY$, with transition probabilities specified by the conditional distribution $W(y|x)$.

For transmission with the rate $R$, a block length-$n$ codebook $\mathcal{C}$ consists of $2^{nR}$ vectors $\mathbf{x}(m) = (x^{(1)}(m),\cdots,x^{(n)}(m))\in\mcX^{n}$. The encoder maps a message $m$, uniformly randomly selected from the set $\mathcal{M} = \{1,\cdots,2^{nR}\}$, to the corresponding codeword $\mathbf{x}(m)$. Upon receiving the output sequence $\mathbf{y} = (y^{(1)},\cdots,y^{(n)})$, the decoder forms an estimate of the message $m$ according to the decoding rule
\begin{equation*}
    \hat{m} = \argmin_{j\in\mathcal{M}}\sum_{k=1}^{n}d(x^{(k)}(j),y^{(k)}),
\end{equation*}
where $d : \mcX\times\mcY\to\mbbR$ is a non-negative function called the \textit{decoding metric}.

With notations defined above, the LM rate, which serves as an achievable lower bound of the mismatch capacity, is defined by the following optimization \cite{Kanabar10409281}
\begin{subequations}\label{eq:IM}
\begin{align}
I_{\mathrm{LM}}(X; Y) \triangleq \min_{{Q}_{X Y} \in \mathcal{P}(\mathcal{X} \times \mathcal{Y})}  & D( {Q}_{X Y} \| P_{X}P_{Y}) \label{IM_obj} \\
\text{s.t.}~\int_{\mcY}{Q}_{X Y}(x, y) \mrmd y & = P_{X}(x),~ \forall x\in \mcX,  \label{IM_Px} \\
\int_{\mcX}{Q}_{X Y}(x, y) \mrmd x & = P_{Y}(y),~ \forall y\in \mcY,  \label{IM_PY} \\
\mathbb{E}_{{Q}_{X Y}}[d(X, Y)] & \leq \mathbb{E}_{P_{X Y}}[d(X, Y)].\label{IM_ineq}
\end{align}
\end{subequations}
Here, $D(\cdot\|\cdot)$ denotes the Kullback–Leibler (KL) divergence,  and
$\mathcal{P}(\mcX\times\mcY)$ denotes the set of all joint probability distributions on $\mcX\times\mcY$.
%

\subsection{Optimal Transport Theory}
The optimal transport theory \cite{OT1941, OT1942}, first proposed by the French mathematician Gaspard Monge \cite{monge1781memoire}, characterizes the minimum cost for transporting one pile of substance to a designated location. 
A central concept within OT theory is the Wasserstein metric \cite{villani2003}, which provides a strong theoretical foundation to measure the difference between two probability distributions through the optimal rearranging cost. 
In particular, the Wasserstein distance is often expressed by the Kantorovich’s formulation, described as
\begin{equation}\label{LM_OT_1}
    W(u,v) = \inf_{\gamma(x, y) \in \Pi(u, v)} \int_{\Omega\times\Omega}d(x,y)\gamma(x, y)\mrmd x \mrmd y,
\end{equation}
where $\Pi(u, v)$ denotes joint probability measures with the marginal measure $u(x)$ and $v(y)$, respectively.

The Wasserstein metric has been found wide applicability across various fields, e.g., the
computer vision \cite{Rub1998Image},
image processing \cite{EM2000},
inverse problem \cite{Earth2018},
and
machine learning \cite{WGAN2021}.
Different numerical methods have been proposed for computing it, including linear programming approaches \cite{linearly2021}, methods based on solving the Monge–Amp\'ere equation \cite{MA2014}, and some combinatorial algorithms \cite{Fili2015}.

Among these, one of the most widely adopted numerical methods is the Sinkhorn algorithm, initially proposed in \cite{sinkhorn1967diagonal} and extensively developed in recent years \cite{2013sinkhorn}. 
To address the high computational complexity of solving the classical OT problem directly, the Sinkhorn algorithm utilize an entropy-regularized version \cite{peyre2019computational} below
\begin{equation}\label{LM_OT_2}
\begin{aligned}
    & W_{\varepsilon}(u,v) = \inf_{\gamma(x, y) \in \Pi(u, v)} \\
    & \left(
    \int_{\Omega\times\Omega}
    \left( d(x,y)\gamma(x, y) + \varepsilon\gamma(x, y)\ln(\gamma(x, y)) \right)
    \mrmd x \mrmd y
    \right),
\end{aligned}  
\end{equation}
where $d(x,y)$ represents the distance.
The optimization \eqref{LM_OT_2} is strictly convex, and has a unique optimal solution.
The Sinkhorn algorithm exploits the structure of this regularized problem and performs iterative updates via alternating solutions, efficiently converging to an approximate solution of the original OT problem while benefiting from improved numerical stability and scalability.

In this work, we observe a structural similarity between the LM rate formulation in \eqref{eq:IM} and the entropy-regularized optimal transport problem in \eqref{LM_OT_2}. 
This observation motivates us to reformulate the LM rate into an equivalent optimal transport framework, enabling the application of a Sinkhorn-type algorithm for efficient numerical solutions.
Details of the reformulation and the proposed algorithm will be presented in subsequent sections.

\section{The LM Rate in OT Formulation}  \label{sec:lmot}
%
We elucidate the connection between the LM rate problem and the entropy-regularized optimal transport.
%
By introducing a Lagrange multiplier $\lambda$ associated with the inequality constraint \eqref{IM_ineq}, the LM rate expression in \eqref{eq:IM} can be reformulated into a structure similar with \eqref{LM_OT_2}, as follows
\begin{equation}\label{LM_T}
\begin{aligned}
    & G_{\lambda}(P_X, P_Y) = \min_{{Q}_{X Y}(x, y) \in \Pi(P_X, P_Y)} \\
    & \left(
    D({Q}_{X Y} \| P_X P_Y) +\lambda \int d(x,y){Q}_{X Y}(x, y)\mrmd x \mrmd y  \right) ,
\end{aligned}
\end{equation}
where ${Q}_{X Y}(x, y) \in \Pi(P_X, P_Y)$ denotes the probability measure with the marginal measure $P_X$ and $P_Y$ in order.

Subsequently, consider a discrete input probability distribution $P_X(x)$ defined on $\mcX$, satisfies 
an average power constraint
\begin{equation}\label{pc}
     \mathbb{E}_{P_X} \|X\|^2 = 1.
\end{equation}
For a given channel transition law $W(y|x)$ and decoding metric $d(x,y)$, the right-hand side of the constraint \eqref{IM_ineq} becomes a constant, defined as
%
    $T = \mathbb{E}_{P_{X Y}}[d(X, Y)].$
%
Furthermore, the relative entropy term in the objective function of \eqref{eq:IM} can be decomposed as
\begin{align*}
    D({Q}_{X Y}\|P_{X}P_{Y}) &= \mathbb{E}_{{Q}_{X Y}} \left(\log {Q}_{X Y}\right) - C, \\
    C &= \mathbb{E}_{P_X}\left(\log P_X \right) + \mathbb{E}_{P_Y}\left(\log P_Y \right).
\end{align*}
Therefore, minimizing the objective function of the LM rate $D({Q}_{X Y}\|P_{X}P_{Y})$ is equivalent to maximizing an entropy term
%
    $-\int_{\mcX}\int_{\mcY}{Q}_{X Y}(x, y)
    \log{Q}_{X Y}(x, y)\mrmd x\mrmd y.$
%

In summary, the formulation of the LM rate in \eqref{IM_obj}-\eqref{IM_ineq} corresponds to an optimal transport problem with 
an additional constraint on the expected decoding cost \eqref{LM_ineq}
\begin{subequations} \label{LM}
\begin{align} 
    \min_{{Q}_{X Y} \in \mathcal{P}(\mathcal{X} \times \mathcal{Y})}
    &
    \int_{\mcX}\int_{\mcY}{Q}_{X Y}(x, y)\log{Q}_{X Y}(x, y)\mrmd x \mrmd y \label{LM_obj} \\
    \text{s.t.}~&\int_{\mcY}{Q}_{X Y}(x, y) \mrmd y= P_{X}(x),\quad \forall x\in \mcX, \label{LM_x} \\
    &\int_{\mcX}{Q}_{X Y}(x, y) \mrmd x = P_{Y}(y),\quad \forall y\in \mcY, \label{LM_y} \\
    &\int_{\mcX}\int_{\mcY}{Q}_{X Y}(x, y) d(x,y)\mrmd x\mrmd y \leq T. \label{LM_ineq}
\end{align}
\end{subequations}
%

Actually, \eqref{LM} shares formal similarities to the entropy-regularized optimal transport problem (EOT) \cite{peyre2019computational}.
The entropy regularization term \eqref{LM_obj} and the extra constraint \eqref{LM_ineq} correspond to the objective function of the EOT problem \eqref{LM_OT_2}.
Furthermore, the constraint \eqref{LM_ineq} imposes an upper bound on the expected distortion under the transport plan ${Q}_{X Y}(x, y)$, which can be interpreted as an upper bound on transport capacity. 
As such, the LM rate optimization \eqref{LM} can also be viewed as a form of 
the capacity-constrained optimal transport \cite{Korman2012OptimalTW, Benamou2014IterativeBP, Wu2022TheDR}. 
This connection 
facilitates the use of the Sinkhorn algorithm \cite{2013sinkhorn} to efficiently compute the LM rate, as discussed in the subsequent section.

\section{The Sinkhorn-type Algorithm} \label{sec:sinkhorn}
In this section, we apply the Sinkhorn algorithm to solve the OT formulation of the LM rate. To this end, we discretize integrals in \eqref{LM} by selecting a uniform grid $\{{y_j}\}_{j=1}^N$ over $\mcY$ and using rectangular integration, which yields the discrete form 
\begin{subequations}\label{OT}
\begin{align}
    \min_{Q_{ij}}\quad & \sum_{i=1}^{M}\sum_{j=1}^{N}Q_{ij}\log Q_{ij} \label{OT_obj} \\
    \text{s.t.}\quad&\sum_{j=1}^{N}Q_{ij} = P_X(x_i) ,\quad i = 1,\cdots,M, \label{OT_Px} \\
    &\sum_{i=1}^{M}Q_{ij} = P_Y(y_j),\quad j = 1,\cdots,N, \label{OT_Py} \\
    & \sum_{i=1}^{M}\sum_{j=1}^{N} d_{ij}Q_{ij}\le T. \label{OT_ineq}
\end{align}
\end{subequations}
%

%
Introducing the dual variables $\bda\in\mbbR^{M}, \bdb\in\mbbR^{N}$ and $\lambda \in\mbbR^{+}$, the Lagrangian of \eqref{OT} is given by

\vspace{-.1in}
\begin{small}
\begin{equation}\label{Lagrangian}
\begin{aligned}
    \mathcal{L}(\bm{Q}; \bda, \bdb, \lambda) \!=\! \sum_{i=1}^{M}\sum_{j=1}^{N}Q_{ij}\log Q_{ij}     
    \!+\!
    \lambda\left(\sum_{i=1}^{M}\sum_{j=1}^{N}d_{ij}Q_{ij} - T\right) &
    \\
    +\sum_{i=1}^{M}\alpha_{i}\left(\sum_{j=1}^{N}Q_{ij} - P_X(x_i)\right) + \sum_{j=1}^{N}\beta_{j}\left(\sum_{i=1}^{M}Q_{ij} - P_Y(y_j)\right) &
    . 
\end{aligned}
\end{equation}
\end{small}

Taking the derivative of $\mathcal{L}(\bm{Q}; \bda, \bdb, \lambda)$ with respect to $Q_{ij}$ leads to 
\begin{equation}\label{gamma}
    {Q}_{ij}^* = \phi_{i}\Lambda_{ij}\psi_{j},
\end{equation}
in which
\begin{equation}\label{dual_var}
    \phi_{i} = \mrme^{-\alpha_{i}-1/2},\;
    \psi_{j} = \mrme^{-\beta_{j}-1/2},\;
    \Lambda_{ij} = \mrme^{-\lambda d_{ij}}.
\end{equation}
Here the bijection exist between $\phi$ and $\alpha$, and between $\psi$ and $\beta$, respectively.

Substituting the above formula into \eqref{OT_Px} and \eqref{OT_Py} yields
\begin{equation*}
\begin{aligned}
    &\phi_{i}\sum_{j=1}^{N}\Lambda_{ij}\psi_{j} = P_X(x_i),~i = 1,\cdots,M,   \\
    &\psi_{j}\sum_{i=1}^{M}\Lambda_{ij}\phi_{i} = P_Y(y_j),~j = 1,\cdots,N.  
\end{aligned}
\end{equation*}
Since $\Lambda_{ij}>0$, we can alternatively update $\phi_i$ and $\psi_j$ according to
\begin{subequations}\label{sink_iter}
\begin{align}
    &\phi_{i}^{\ell+1} = P_X(x_i)/\sum_{j=1}^{N}\Lambda_{ij}^{\ell}\psi_{j}^{\ell},~~ i = 1,\cdots,M, \label{sink_px} \\
    &\psi_{j}^{\ell+1} = P_Y(y_j)/\sum_{i=1}^{M}\Lambda_{ij}^{\ell}\phi_{i}^{\ell+1},~ j = 1,\cdots,N. \label{sink_py}
\end{align}
\end{subequations}
This iterative formula \eqref{sink_iter} is the well-known Sinkhorn algorithm \cite{sinkhorn1967diagonal}. 

Different from the classical OT problem, we need to handle the extra constraint \eqref{OT_ineq}. Taking the derivative of the Lagrangian $\mathcal{L}({Q}_{X Y}; \bm\phi, \bm\psi, \lambda)$ with respect to $\lambda$ yields
\begin{equation}\label{G_lambda}
    F(\lambda; \bm\phi, \bm\psi)\triangleq \sum_{i=1}^{M}\sum_{j=1}^{N} \phi_{i}\psi_{j} d_{ij} \mrme^{-\lambda d_{ij}} - T.
\end{equation}
Noticing that
\begin{equation*}
    F^{'}(\lambda; \bm\phi, \bm\psi)= -\sum_{i=1}^{M}\sum_{j=1}^{N}\phi_{i}\psi_{j}d_{ij}^{2}\mrme^{-\lambda d_{ij}} < 0,
\end{equation*}
the function $F(\lambda; \bm\phi, \bm\psi)$ is monotonically decreasing. 
Since the variable $\lambda$ is restricted to the interval $[0, +\infty)$ and its gradient is valid, the gradient projection method provides an effective iterative framework for updating $\lambda$, \thatis,
$$\lambda^{\ell + 1} = \left [\lambda^{\ell} + \tau F(\lambda^{\ell}; \bm\phi, \bm\psi) \right]_{+},$$ where $\tau$ is the step size.

Finally, the pseudo-code of the Sinkhorn-type algorithm is summarized in Algorithm \ref{alg:OT}.

\begin{algorithm}[ht]
\caption{The Sinkhorn-type Algorithm}
\label{alg:OT}
\begin{algorithmic}[1]
    \REQUIRE Decoding metric $d_{ij}$; Marginal distributions $P_X(x_i)$, $P_Y(y_j)$; Iteration number $L_0$, Step size $\tau$.
    \ENSURE Minimal value of $\sum_{i=1}^{M}\sum_{j=1}^{N}Q_{ij}\log Q_{ij}$.
    \STATE \textbf{Initialization:} $\bm{\phi} = \mathbf{1}_{M}, \bm{\psi} = \mathbf{1}_{N}, \lambda = 1$;
    \FOR{$\ell = 1 : L_0$}
    \STATE $\Lambda_{ij} \gets \mrme^{-\lambda d_{ij}}$,\quad $i = 1,\cdots,M,\ j = 1,\cdots,N$
    \FOR{$i = 1 : M$}
    \STATE $\phi_{i} \gets P_X(x_i)/\sum_{j=1}^{N}\Lambda_{ij}\psi_{j}$
    \ENDFOR
    \FOR{$j = 1 : N$}
    \STATE $\psi_{j} \gets P_Y(y_j)/\sum_{i=1}^{M}\Lambda_{ij}\phi_{i}$
    \ENDFOR
    \STATE 
    $\lambda \leftarrow \left[\lambda + \tau{F(\lambda; \bm\phi, \bm\psi)}\right]_{+}$
    \ENDFOR
    \RETURN $\sum_{i=1}^{M}\sum_{j=1}^{N} \phi_{i}\psi_{j}\Lambda_{ij}\log\left(\phi_{i}\psi_{j} \Lambda_{ij}\right)$
\end{algorithmic}
\end{algorithm}
%
%

\section{Convergence of the Proposed Sinkhorn-type Algorithm} \label{sec:analy}
\subsection{Dual Form of the LM Rate}
%
The LM rate \eqref{eq:IM} is a convex optimization problem, which forms the basis for the subsequent convergence analysis. 
For clarity, we present the matrix formulation of \eqref{gamma} as follows:
\begin{equation}\label{eq:bdga}
    \mathcal{Q}(\bda,\bdb,\lambda)
    \!=\! \operatorname{Diag}\left(\mathrm{e}^{-\bda-\mathbf{1}_{M}/2}\right) \bm\Lambda \operatorname{Diag}\left(\mathrm{e}^{-\bdb-\mathbf{1}_{N}/2}\right).
\end{equation}
The elements of \eqref{eq:bdga} are $\mathcal{Q}(\alpha_i, \beta_j, \lambda)$ defined in \eqref{gamma}. 

Based on the derivation outlined above, we introduce a novel dual form of the LM rate, which incorporates the coherent marginal distributions as parameters and features a kernel matrix multiplication structure. 
Under a prescribed input distribution $P_X$ and a channel transition law $W(y|x)$, the dual form of the LM rate \eqref{eq:IM} is given by
\begin{equation}\label{dual_new_0}
\begin{aligned}
    I_{\mathrm{LM}}(X, Y) =  \max_{\substack{\bda,\bdb, \\ \lambda > 0}}   
    \left( H(X) + H(Y) - \left\langle \bda, P_X(\bdx) \right\rangle
    \right. & \\
    \left. 
    - \left\langle \bdb, P_Y(\bdy) \right\rangle
    - \left\langle \lambda, T \right\rangle
    - \mathbf{1}_{M}^{\top} \mathcal{Q}(\bda,\bdb,\lambda) \mathbf{1}_{N}
    \right) & ,
\end{aligned}
\end{equation}
where $H(\cdot)$ denotes the information entropy.
The objective function in \eqref{dual_new_0} is obtained by substituting dual variables \eqref{gamma} into the Lagrangian \eqref{Lagrangian}.

\begin{proposition} \label{prop:equi}
    The dual form \eqref{dual_new_0} of the LM rate  is equivalent to \eqref{Scarlett_dual_LM_1} below, 
    which was introduced in \cite[p. 19]{2020Information}.
\begin{equation} \label{Scarlett_dual_LM_1}
    \max _{\substack{\zeta \geq 0, \\a(\cdot)}} \sum_{x, y} P_X(x) W(y \mid x) \log \frac{q(x, y)^{\zeta} e^{a(x)}}{\sum_{\bar{x}} P_X(\bar{x}) q(\bar{x}, y)^{\zeta} e^{a(\bar{x})}}.
\end{equation}
\end{proposition}

A key distinction lies in the fact that the proposed dual expression \eqref{dual_new_0} is extended from the entropy-regularized optimal transport (OT) model for the LM rate problem,  whereas the dual form \eqref{Scarlett_dual_LM_1} in Scarlett's work \cite{2020Information} is derived via Jensen's inequality. 
Nevertheless, the equivalence of these two expressions is demonstrated in Appendix \ref{app:equi}.
%

\subsection{Property of the Optimal Point}\label{sec:prop}
To discuss our Sinkhorn-type algorithm, we consider the dual problem of \eqref{OT} in its convex minimization form \eqref{dual_problem} for simplicity:
\begin{equation}\label{dual_problem}
\begin{aligned}
    \min_{\substack{\bda,\bdb, \\ \lambda \geq 0}} g(\bda,\bdb,\lambda)
    =\min_{\substack{\bda,\bdb, \\ \lambda \geq 0}} 
    \left(
    \mathbf{1}_{M}^{\top} \mathcal{Q}(\bda,\bdb,\lambda) \mathbf{1}_{N}
    \right. & \\
    \left.
    + \left\langle \bda, P_X(\bdx) \right\rangle + \left\langle \bdb, P_Y(\bdy) \right\rangle 
    + \left\langle \lambda, T \right\rangle
    \right) & .
\end{aligned}
\end{equation}

As discussed in Section \ref{sec:sinkhorn}, the dual variables are updated sequentially via the following update scheme: 
\begin{subequations} \label{update_notation}
\begin{align}
\bda^{\ell+1} &= \argmin_{\bda} g(\bda, \bdb^{\ell}, \lambda^{\ell}), \label{up_bda_no} \\
\bdb^{\ell+1} &= \argmin_{\bdb} g(\bda^{\ell+1}, \bdb, \lambda^{\ell}), \label{up_bdb_no} \\
\lambda^{\ell+1} &= \left[\lambda^{\ell} - \tau \nabla_{\lambda} g(\bda^{\ell+1}, \bdb^{\ell+1}, \lambda^{\ell})\right]_+. \label{up_lam_no}
\end{align}
\end{subequations}
%
%
%
Here the step size $\tau$ can be chosen to satisfy $\tau \leq 1/L_{\lambda}$ for convergence \cite{nocedal2006numerical}, 
where $L_{\lambda}$ is 
the bound of $\nabla_{\lambda} g(\bda^{\ell+1}, \bdb^{\ell+1}, \lambda^{\ell})$ in the update scheme.
Before proceeding with the analysis, we outline the underlying assumptions.
\begin{assp}\label{assp:02}
    The probability measures $P_X$ and $P_Y$ are strictly positive over their respective supports $\mathcal{X}$ and $\mathcal{Y}$, \thatis, the probabilities of all points are non-zero.
\end{assp}


\begin{assp}\label{assp:03}
    The decoding metric $d\colon\mathcal{X}\times\mathcal{Y}\to\mathbb{R}$ is centrally symmetric, i.e., $d(x,y) = d(-x,-y)$ for all $(x,y)\in\mathcal{X}\times\mathcal{Y}$. 
    The alphabets $\mathcal{X}$ and $\mathcal{Y}$ are centrally symmetric sets.
\end{assp}


These assumptions constitute standard conditions applicable to broad scenarios.
%
%
Assumption \ref{assp:02} excludes zero-probability constellation points, as their contribution to mutual information vanishes. This aligns with standard practice in discrete communication systems \cite{2000Mismatched}. 
Assumption \ref{assp:03} is frequently validated across many scenarios. A notable example is the Additive White Gaussian Noise (AWGN) channel, where the decoding metric $d(x, y)$ is selected to be the distance metric.
%


%
%
Meanwhile, the following technical lemma is established to support the subsequent proofs in the theorems.
\begin{lemma}\label{lem:kernel}
Define the matrices $\widetilde{\bm D} \in \mbbR^{MN \times 1}$ and $\bm K \in \mbbR^{MN \times (M+N)}$ as
\begin{equation}\label{eq:kd}
\begin{aligned}
    \widetilde{\bm D} & \!=\! \left( d_{11}, d_{12}, \!\cdots\!, d_{1N}, d_{21}, \!\cdots\!, d_{2N},d_{31}, \cdots, d_{MN}\right)^{\top}, \\    
    \bm K & \!=\! \left( \bm I_{M} \otimes \mathbf{1}_{N}, \,\, \mathbf{1}_{M} \otimes \bm I_{N} \right).
\end{aligned}
\end{equation}
Then the kernel of the matrix $\left( \bm K, \widetilde{\bm D} \right) \in \mbbR^{MN \times (M+N+1)}$ has the formulation
%
\begin{equation}\label{kernel}
    \operatorname{ker}\left[ \left( \bm K, \widetilde{\bm D} \right) \right] 
    = \left\{ s\cdot(\mathbf{1}_{M}; -\mathbf{1}_{N}; 0)
    \mid  s\in\mbbR\right\}.
\end{equation}
\end{lemma}

The proof of Lemma \ref{lem:kernel} is shown in Appendix \ref{app:kernel}.

\begin{remark}
    A key condition utilized in the proof of Lemma \ref{lem:kernel} is that the reshaped decoding metric $\widetilde{\bm{D}}$ cannot be expressed as a linear combination of column vectors in the coefficient matrix $\bm{K}$. 
    Intuitively, this condition is reasonable and can be satisfied under certain restrictions.
    Assumption \ref{assp:03} serves as a constraint that ensures the validity of this conditions. 
\end{remark}

Now, we proceed to demonstrate the theorem concerning the optimal point.

\begin{theorems}\label{thm:01}
The Karush-Kuhn-Tucker point of the optimization problem \eqref{OT} corresponds to the optimal solution
$(\bda^{*},\bdb^{*},\lambda^{*})$ of \eqref{dual_problem},
where the dual variable $\lambda^{*}$ and its corresponding optimal point are unique.
%
\end{theorems}

\begin{proofs}
The Karush-Kuhn-Tucker (KKT) conditions of the optimization \eqref{OT} are
\begin{subequations}
\begin{align}
    &Q_{ij}  = \mrme^{-\alpha_i-\beta_j-\lambda d_{ij} - 1}, ~\forall i,j,
    \label{eq:Stationarity}\\
    & \sum_{j = 1}^{N} Q_{ij}  = P_X(x_i),~~ 
    \sum_{i = 1}^{M} Q_{ij} = P_Y(y_j),~ \forall i,j,
    \label{eq:Primal}\\
    & \sum_{j=1}^{N} d_{ij}Q_{ij} - T  \leq 0, \label{eq:pr}\\
    & \lambda \geq  0,~\lambda \left(  \sum_{i=1}^{M}\sum_{j=1}^{N}d_{ij}Q_{ij} - T \right) = 0.~  \label{eq:du} 
\end{align}
\end{subequations}

Substituting \eqref{eq:Stationarity} into \eqref{eq:Primal}, the KKT point of the optimization problem \eqref{OT}, denoted by $(\bda^{*},\bdb^{*},\lambda^{*})$ satisfies the first-order condition $\nabla_{\bda, \bdb}~ g(\bda^{*},\bdb^{*}, \lambda^{*})=\bm{0}$ as derived from equality constraints. 

We proceed to analyze $\lambda^{*}$ under two distinct cases.
First, the constraint \eqref{OT_ineq} (\thatis, \eqref{eq:pr}) is an inactive constraint, which leads to $\lambda^{*} = 0$ according to \eqref{eq:du}.
%
%

Second, considering the case when the constraint \eqref{OT_ineq} is an active constraint,
we have $\nabla_{\lambda} g(\bda^{*},\bdb^{*},\lambda^{*})=\bm{0}$.
In the following, we analyze the second-order condition.
%
%
The gradient $g(\bda, \bdb, \lambda)$ is written as follows

\vspace{-.1in}
\begin{small}
\begin{equation*}
\nabla g(\bda, \bdb,\lambda) \!=\! \left(
\begin{array}{l}
    P_X(\bdx)-\operatorname{Diag}\left(\mathrm{e}^{-\bda-\mathbf{1}_{M}/2}\right) \bm\Lambda \,\,\mathrm{e}^{-\bdb-\mathbf{1}_{N}/2} \\
    P_Y(\bdy)-\operatorname{Diag}\left(\mathrm{e}^{-\bdb-\mathbf{1}_{N}/2}\right) \bm\Lambda^{\top} \mathrm{e}^{-\bda-\mathbf{1}_{M}/2}\\
    T-\left(\mathrm{e}^{-\bda-\mathbf{1}_{M}/2}\right)^{\top} (\bm D\odot \bm\Lambda) \,\,\mathrm{e}^{-\bdb-\mathbf{1}_{N}/2}
\end{array}
\right),
\end{equation*}
\end{small}

\noindent
where $\bm D = \left(d_{ij}\right)_{M\times N}$ is the decoding metric.
And the Hessian of $g(\bda, \bdb, \lambda)$ 
is given by

\vspace{-.1in}
\begin{small}
\begin{equation*}
\begin{aligned}
    & H_g(\bda, \bdb, \lambda) =  \\
    & \left[
\begin{array}{ccc}
    \operatorname{Diag}\left(\mathcal{Q} \mathbf{1}_{N}\right) & \mathcal{Q} & (\bdD\odot \mathcal{Q})\mathbf{1}_{N}\\
    \mathcal{Q}^{\top} & \operatorname{Diag}\left(\mathcal{Q}^{\top} \mathbf{1}_{M}\right) & (\bdD\odot \mathcal{Q})^{\top} \mathbf{1}_{M} \\
    ((\bdD\odot \mathcal{Q})\mathbf{1}_{N})^{\top} & \mathbf{1}_{M}^{\top}(\bdD\odot \mathcal{Q}) & \mathbf{1}_{M}^{\top} (\bdD\odot\bdD\odot \mathcal{Q}) \mathbf{1}_{N}
\end{array}
    \right].
\end{aligned}
\end{equation*}
\end{small}
%

%
 For arbitrary $\left(\bm a^{\top}, \bm b^{\top}, c\right)\in(\mbbR^{1\times M},\mbbR^{1\times N},\mbbR)$, we obtain the following 
\begin{equation}\label{kernel_inequalty}
\begin{aligned}
    & \left(\bm a^{\top}, \bm b^{\top}, c\right) 
    H_g(\bda, \bdb, \lambda)
    \left(\bm a^{\top}, \bm b^{\top}, c\right)^{\top} \\
    & = \sum_{i = 1}^{M} \sum_{j = 1}^{N} \mathcal{Q}(\alpha_{i},\beta_{j},\lambda)\left(a_i+b_j+d_{ij}c\right)^2 \geq 0,
\end{aligned}
\end{equation}
which suggests that $H_g(\bda, \bdb, \lambda)$ is a positive semi-definite matrix.
Thus, the kernel is given by
\begin{equation*} \label{eq:kernel}
\begin{aligned}
    & \operatorname{ker}\left[H_g(\bda, \bdb, \lambda)\right] \\
    & = \left\{
    \left(\bm a; \bm b; c\right) \in \mathbb{R}^{(M+N+1)\times 1}  \mid 
    a_i+b_j+d_{ij}c=0, ~ \forall i,j \right\} \\
    & \triangleq \left\{\left(\bm a; \bm b; c\right) \in \mathbb{R}^{(M+N+1)\times 1} \mid 
    \left(\bm K, \widetilde{\bm D}\right)
    \left(\bm a; \bm b; c\right)
    = \bm{0} \right\}.
\end{aligned}
\end{equation*}
Here $\bm K \in \mbbR^{MN \times (M+N)}$ is the coefficient matrix corresponding to $\bm a$ and $\bm b$, and 
$\widetilde{\bm D}\in \mbbR^{MN \times 1}$ is the coefficient matrix corresponding to $c$, which is the same as \eqref{eq:kd}.
Lemma \ref{lem:kernel} indicates that the kernel is equivalent to
%
\begin{equation}\label{kernel_2}
    \operatorname{ker}\left[ H_g(\bda, \bdb, \lambda) \right] 
    = \left\{ s\cdot(\mathbf{1}_{M}; -\mathbf{1}_{N}; 0)\
    \mid  s\in\mbbR\right\}.
\end{equation}

Hence, $H_g(\bda, \bdb, \lambda)$ is strictly positive definite on this subspace of $\lambda >0$ as indicated by \eqref{kernel_2}. 
Then the KKT point $(\bda^{*},\bdb^{*},\lambda^{*})$ meets the conditions
$$
\nabla_{\bda, \bdb, \lambda} g(\bda^{*},\bdb^{*},\lambda^{*})=\bm{0}, \quad \nabla^2 g (\bda^{*},\bdb^{*},\lambda^{*}) \succ 0,
$$
which suggests that the KKT point is an optimal point.

We also claim that the optimal solution of the primary problem \eqref{OT}, as deduced from $(\bda^{*},\bdb^{*},\lambda^{*})$ is unique, \thatis, 
$$ Q_{ij}^{*} = \exp(-\alpha_{i}^{*}-1/2)\exp(-\lambda^{*} d_{ij})\exp(-\beta_{j}^{*}-1/2) $$ 
is unique.
To argue by contradiction, assume that there exists a different $Q_{ij}^{\prime} = Q_{ij}^{*}$, which corresponds to two distinct sequences of dual variables $(\bda^{\prime},\bdb^{\prime},\lambda^{\prime})$ and $(\bda^{*},\bdb^{*},\lambda^{*})$, respectively.
This yields an equation system between the dual variables, described by
\begin{equation}\label{eq:unique}
    \left(\alpha_i^{\prime} - \alpha_i^{*} \right) + 
    \left(\beta_j^{\prime} - \beta_j^{*} \right) + 
    d_{ij}\left( \lambda^{\prime} - \lambda^{*} \right) = 0, ~ \forall i, j.
\end{equation}
Similar to Lemma \ref{lem:kernel}, the above equation system \eqref{eq:unique} does not have any solution when $\lambda^{\prime} \neq \lambda^{*}$,
leading to a contradiction.
Therefore, we conclude that the optimal solution of the primal problem \eqref{OT}, denoted by $Q_{ij}^{*}$, and its corresponding dual variable $\lambda^*$ are unique.
%
\end{proofs}
%

%
\begin{remark}\label{rmk:finite}
    From the dual structure \eqref{dual_problem}, multiplier $\lambda$ can be roughly interpreted as corresponding to the slope of a tangent line on the function curve of the achievable rate $I_{\mathrm{LM}}$ with respect to a threshold $T$, similar to previous discussion in the rate-distortion problem \cite{wu2022communication}. 
    Therefore, in the proof of Theorem \ref{thm:01}, we implicitly assume that there exists a finite optimal point $\lambda^*$.
\end{remark}
%

\subsection{Analysis of the Convergence Rate}
%
This subsection derives the convergence rate and termination criteria for the Sinkhorn-type algorithm.
Define the bound on the error function after $\ell$-th iteration as
\begin{equation}
    e^{\ell}(\bda^{\ell},\bdb^{\ell},\lambda^{\ell}) \triangleq g(\bda^{\ell},\bdb^{\ell},\lambda^{\ell}) - g(\bda^{*},\bdb^{*},\lambda^{*}).
\end{equation}
The main result is summarized in Theorem \ref{thm:main}, which implies that Algorithm \ref{alg:OT} has the sublinear rate of convergence.

\begin{theorems}\label{thm:main}
Algorithm \ref{alg:OT} outputs a solution $\mathcal{Q}(\bda^{\ell},\bdb^{\ell},\lambda^{\ell})$ satisfying the error bound
\begin{equation*}
    e^{\ell}(\bda^{\ell},\bdb^{\ell},\lambda^{\ell}) \leq \varepsilon
\end{equation*}
when the number of iteration $\ell$ meets the condition
\begin{equation}
    \ell \leq \Big(\frac{1}{\varepsilon} - \frac{1}{e^{0}}\Big) \cdot \frac{1}{8S_{0}^2(1+L_{\lambda})},
\end{equation}
where $e^{0} \triangleq e(\bda^{0},\bdb^{0},\lambda^{0})$.
\end{theorems}
Before estimating the convergence rate, we need some lemmas that will be invoked sequentially. 
Lemma \ref{lem:lips} verifies 
a Lipschitz-type inequality in the $\lambda$-direction.
Lemma \ref{lem:grad} exploits this condition to bound the objective and gradient decrease during the $\lambda$-update. 
Lemma \ref{lem:bound} then derives uniform component-wise bounds for $\bda^{\ell},\bdb^{\ell},\lambda^{\ell}$. 
Finally, Lemma \ref{lem:decr} leverages these conditions to quantify the objective and gradient decrease in the $\beta$-direction.

%
\begin{lemma}\label{lem:lips}
The dual objective function $g(\bda,\bdb,\lambda)$ satisfies the condition
%
\begin{equation}\label{L-smooth}
\begin{aligned}
    & g(\bda^{\ell+1}, \bdb^{\ell+1}, \lambda^{\ell+1}) \leq g(\bda^{\ell+1}, \bdb^{\ell+1}, \lambda^{\ell})  \\
    & + \langle \nabla_{\lambda}g(\bda^{\ell+1}, \bdb^{\ell+1}, \lambda^{\ell}), \lambda^{\ell+1}-\lambda^{\ell} \rangle+\frac{L_{\lambda}}{2}\|\lambda^{\ell+1}-\lambda^{\ell}\|^{2}_{2},
\end{aligned}
\end{equation}
where $L_\lambda$ is a finite constant. 
%
\end{lemma}

The proof of Lemma \ref{lem:lips} is provided in Appendix \ref{app:lips}.
%

%
Further, Lemma \ref{lem:grad} gives estimates for the reduction value of the objective function $g$ and the gradient at the direction $\lambda$ based on the property of $L$-smooth.
\begin{lemma}\label{lem:grad}
For the $\ell$-th iteration of $\lambda$, the decrease of the objective function is bounded by the distance of two consecutive iteration points 
    \begin{equation}\label{esti_lam}
    \begin{aligned}
        & g(\bda^{\ell+1}, \bdb^{\ell+1},\! \lambda^{\ell}) \!-\! g(\bda^{\ell+1}, \bdb^{\ell+1}, \!\lambda^{\ell+1}) \!\geq \!\frac{L_{\lambda}}{2}\!\|\lambda^{\ell}\!-\!\lambda^{\ell+1}\|^{2}_{2}.
    \end{aligned}
    \end{equation}
    Also, its gradient can be estimated by
    \begin{equation}\label{gddd_lam}
    \begin{aligned}
        & \left \langle \nabla_{\lambda} g(\bda^{\ell+1}, \bdb^{\ell}, \lambda^{\ell}), \lambda^{\ell} - \lambda^{*} \right \rangle  
        \\
        &\quad \leq 2M_{0}L_{\lambda} \|\lambda^{\ell} - \lambda^{\ell+1}\|_{2} \!+\!  M_{0}\|\nabla_{\bdb} g(\bda^{\ell+1}, \bdb^{\ell}, \lambda^{\ell})\|_{1},
    \end{aligned}
    \end{equation}
    where $M_{0}$ is a finite constant. 
\end{lemma}

The proof of Lemma \ref{lem:grad} is shown in Appendix \ref{app:grad}. 

In the following, we examine Sinkhorn iterations with respect to variables $\bda$ and $\bdb$.
For all iterative points, Lemma \ref{lem:bound} provides a unified bound on components of dual variables $\bda^{\ell},\bdb^{\ell},\lambda^{\ell}$, following a similar approach to Dvurechensky$'$s work \cite[Lemma 1]{2018Dvurechensky}).
\begin{lemma} \label{lem:bound}
For the iteration number $\ell\geq 0$,
the dual variables $\bda^{\ell}, \bdb^{\ell}, \lambda^{\ell}$ can be bounded by a finite constant $S_{0}$, \thatis,
\begin{subequations}
\begin{align}
    \max_{i}\alpha^{\ell}_{i}-\min_{i}\alpha^{\ell}_{i}\leq S_{0},& \quad \max_{j}\beta^{\ell}_{j}-\min_{j}\beta^{\ell}_{j}\leq S_{0}, \\
    \max_{i}\alpha^{*}_{i}-\min_{i}\alpha^{*}_{i}\leq S_{0},& \quad \max_{j}\beta^{*}_{j}-\min_{j}\beta^{*}_{j}\leq S_{0}, 
    \\
    \sup_{\ell} \vert\lambda^{\ell}-\lambda^{*}\vert \leq S_{0}, & \quad M_{0} \leq S_{0}. 
    \label{eq:bon}
\end{align}
\end{subequations}
\end{lemma}

The last inequality is introduced for notational convenience.
The proof of Lemma \ref{lem:bound} is provided in Appendix \ref{app:bound}.

The Sinkhorn iterations generate explicit relationships between iterative variables, which provide a foundation for analyzing the convergence rate in subsequent proofs.
The derivation with respect to $\bda, \bdb$ in each iteration yields the following equations about gradients
\begin{equation}\label{grad_equation}
\begin{aligned}   
    \nabla_{\bda}g(\bda, \bdb^{\ell}, \lambda^{\ell}) &=  P_X(\bdx) - \mathcal{Q}(\bda, \bdb^{\ell}, \lambda^{\ell})\mathbf{1}_{N},
    \\
    \nabla_{\bdb}g(\bda^{\ell+1}, \bdb, \lambda^{\ell}) &=  P_Y(\bdy) - \mathcal{Q}(\bda^{\ell+1}, \bdb, \lambda^{\ell})^{\top}\mathbf{1}_{M} .
\end{aligned}
\end{equation}
Combining these with \eqref{update_notation}, the alternating update in each iteration can be written as
\begin{equation}\label{updating_ab}
\begin{aligned}
\bda^{\ell+1}-\bda^{\ell} & =-\ln {P_X(\bdx)}+\ln (\mathcal{Q}(\bda^{\ell}, \bdb^{\ell}, \lambda^{\ell}) \mathbf{1}_{N}), \\
\bdb^{\ell+1}-\bdb^{\ell} & =-\ln {P_Y(\bdy)}+\ln (\mathcal{Q}(\bda^{\ell+1}, \bdb^{\ell}, \lambda^{\ell})^{\top} \mathbf{1}_{M}) .
\end{aligned}
\end{equation}
%

Building upon this, Lemma \ref{lem:decr} provides a uniform lower bound on the objective decrease between consecutive iterates and a matching upper bound for the gradient along the $\bdb$-direction.

\begin{lemma}\label{lem:decr}
    For the $\ell$-th iteration of $\bdb$, the decrease of the objective function is bounded by the gradient
    \begin{equation}\label{esti_beta}
    \begin{aligned}
        g(\bda^{\ell+1}, \bdb^{\ell}, \lambda^{\ell}) - g(\bda^{\ell+1}, \bdb^{\ell+1}, \lambda^{\ell}) &
        \\
        \geq \frac{1}{2}\|\nabla_{\bdb}g(\bda^{\ell+1}, \bdb^{\ell}, \lambda^{\ell})\|_{1}^2 & ,
    \end{aligned}
    \end{equation}
    And its gradient can be estimated by
    \begin{equation}\label{gddd_beta}
    \begin{aligned}
        \left \langle \nabla_{\bdb} g(\bda^{\ell+1}, \bdb^{\ell}, \lambda^{\ell}), \bdb^{\ell} - \bdb^{*} \right \rangle  
        & \!\leq\! S_{0} \|\nabla_{\bdb} g(\bda^{\ell+1}, \bdb^{\ell}, \lambda^{\ell})\|_{1}.
    \end{aligned}
    \end{equation}
\end{lemma}

The proof of Lemma \ref{lem:decr} is shown in Appendix \ref{app:decr}. 
%

In accordance with Lemma \ref{lem:grad} and Lemma \ref{lem:decr}, it is possible to estimate the error function $e^{\ell}(\bda^{\ell},\bdb^{\ell},\lambda^{\ell})$.
With the preparatory results in place, we proceed to prove the main results shown in Theorem \ref{thm:main}.

\begin{proofs}
Since the objective function $g(\bda, \bdb, \lambda)$ is convex, we have
\begin{equation}\label{eq:thm:21}
\begin{aligned}
    & e(\bda^{\ell+1},\bdb^{\ell},\lambda^{\ell}) = g(\bda^{\ell+1},\bdb^{\ell},\lambda^{\ell}) - g(\bda^{*},\bdb^{*},\lambda^{*}) \\
    & \quad \leq
    \left \langle \nabla_{\lambda} g(\bda^{\ell+1},\bdb^{\ell},\lambda^{\ell}), \lambda^{\ell} - \lambda^{*} \right \rangle \\
    & \quad \quad + \left \langle \nabla_{\bdb} g(\bda^{\ell+1},\bdb^{\ell},\lambda^{\ell}), \bdb^{\ell} - \bdb^{*} \right \rangle \\
    & \quad \quad +  
    \left \langle \nabla_{\bda} g(\bda^{\ell+1},\bdb^{\ell},\lambda^{\ell}), \bda^{\ell+1} - \bda^{*} \right \rangle \\
    & \quad \leq 2S_{0}L_{\lambda} \|\lambda^{\ell} - \lambda^{\ell+1}\|_{2} +  2S_{0}\|\nabla_{\bdb} g(\bda^{\ell+1}, \bdb^{\ell}, \lambda^{\ell})\|_{1},
\end{aligned}
\end{equation}
where the last inequality holds according to \eqref{gddd_lam}, \!\eqref{eq:bon}, and \eqref{gddd_beta}
and the optimal condition $\nabla_{\bda} g(\bda^{\ell+1},\bdb^{\ell},\lambda^{\ell}) = 0$. 

On the other hand, \eqref{esti_lam} and \eqref{esti_beta} yield
\begin{equation}\label{eq:thm:22}
\begin{aligned}
    & e(\bda^{\ell+1},\bdb^{\ell},\lambda^{\ell})
    - e(\bda^{\ell+1},\bdb^{\ell+1},\lambda^{\ell+1}) \\
    & \quad = g(\bda^{\ell+1},\bdb^{\ell},\lambda^{\ell}) - g(\bda^{\ell+1},\bdb^{\ell+1},\lambda^{\ell+1}) \\
    & \quad \geq
    \frac{1}{2}\|\nabla_{\bdb}g(\bda^{\ell+1}, \bdb^{\ell}, \lambda^{\ell})\|_{1}^2 + \frac{L_{\lambda}}{2}\|\lambda^{\ell+1}-\lambda^{\ell}\|^{2}_{2}
    \\
    & \quad \geq \frac{1}{8S_{0}^2(1+L_{\lambda})} \\
    & ~ \cdot \left(2S_{0}L_{\lambda}\|\lambda^{\ell} - \lambda^{\ell+1}\|_{2} +  2S_{0}\|\nabla_{\bdb} g(\bda^{\ell+1}, \bdb^{\ell}, \lambda^{\ell})\|_{1}\right)^2 \\
    & \quad \geq \frac{1}{8S_{0}^2(1+L_{\lambda})}\left(e(\bda^{\ell+1},\bdb^{\ell},\lambda^{\ell})\right)^2,
\end{aligned}
\end{equation}
where the second inequality holds due to the Cauchy-Schwartz inequality, and the third inequality is based on \eqref{eq:thm:21}.

Upon the division by $e(\bda^{\ell+1},\bdb^{\ell},\lambda^{\ell}) \cdot e(\bda^{\ell+1},\bdb^{\ell+1},\lambda^{\ell+1})$ in both sides of \eqref{eq:thm:22}, the inequality becomes
\begin{equation}\label{eq:thm:23}
\begin{aligned}
    &\frac{1}{e(\bda^{\ell+1},\bdb^{\ell+1},\lambda^{\ell+1})}  \\
    & \quad \geq \frac{1}{e(\bda^{\ell+1},\bdb^{\ell},\lambda^{\ell})}
    + \frac{1}{8S_{0}^2(1+L_{\lambda})}\frac{e(\bda^{\ell+1},\bdb^{\ell},\lambda^{\ell})}{e(\bda^{\ell+1},\bdb^{\ell+1},\lambda^{\ell+1})}\\
    & \quad \geq \frac{1}{e(\bda^{\ell},\bdb^{\ell},\lambda^{\ell})}
    + \frac{1}{8S_{0}^2(1+L_{\lambda})}.
\end{aligned}
\end{equation}
The last inequality holds since
\begin{equation*}
\begin{aligned}
    e(\bda^{\ell},\bdb^{\ell},\lambda^{\ell}) - e(\bda^{\ell+1},\bdb^{\ell},\lambda^{\ell})& \geq 0, \\
    e(\bda^{\ell+1},\bdb^{\ell},\lambda^{\ell}) - e(\bda^{\ell+1},\bdb^{\ell+1},\lambda^{\ell+1})& \geq 0.
\end{aligned}
\end{equation*}

Summing up the inequality \eqref{eq:thm:23} by $\ell$, we obtain
\begin{equation}\label{linear_rate}
    \frac{1}{e(\bda^{\ell+1},\bdb^{\ell+1},\lambda^{\ell+1})} 
    \geq \frac{1}{e^0} + \frac{\ell +1 }{8S_{0}^2(1+L_{\lambda})}.
\end{equation}
Thus, Algorithm \ref{alg:OT} converges with at least a sub-linear rate. 
%
\end{proofs}

\section{Numerical Simulations} \label{sec:numer}
In this section, the OT-based model and Sinkhorn-type algorithm are applied to compute the LM rate for different modulation schemes over additive white Gaussian noise (AWGN) channels, subject to rotation and scaling. 
The channel model is described by the following transition law:
\begin{equation}\label{awgneq}
    Y = H X + Z, \ \text{with } Z\sim \mathcal{N}(0,\sigma_{Z}^{2}).
\end{equation}

The focus is on a finite and discrete channel input alphabet $\mcX$, which leads to the inclusion of truncation and discretization of the channel output alphabet $\mcY$.
Since $\mcY \subset \mbbR^2$, we can truncate 
$\mcY$ by defining a sufficiently large region, for example, the square region $[-8, 8]\times[-8, 8]$, based on reasonable assumptions, e.g. $\eta\le 1,\sigma_z^2\le \frac{1}{2}$. 
The truncated region is then discretized into a set of uniform grid points $\{y_j\}_{j=1}^{N}$ in the following manner:
\begin{equation*}
\begin{aligned}
    y_{r\sqrt{N}+s} &= (-8 + r\Delta y, -8 + (s-1)\Delta y), \Delta y = \frac{16}{\sqrt{N}-1},\\
    r &= 0,1,\cdots,\sqrt{N}-1, s = 1,2,\cdots,\sqrt{N},
\end{aligned}
\end{equation*}
The matrix $H\in\mbbR^{2\times 2}$ in \eqref{awgneq} is a combination of rotation and scaling transformations, given by
\begin{equation*}
    H = \begin{pmatrix} \eta_1 & 0 \\ 0 & \eta_2 \end{pmatrix}\begin{pmatrix} \cos \theta & \sin \theta\\ -\sin \theta & \cos \theta \end{pmatrix}.
\end{equation*}
Here parameters $\eta_1,\;\eta_2$ indicate the scaling of the signal, and the parameter $\theta$ indicates the degree of rotation on the signal. 
In this work, we consider the case where $\eta_1=1$ and $\eta_2=\eta$. 
The specific values of $\eta$ and $\theta$ are unknown in advance.

The decoding metric is defined as $d(x,y) \triangleq \|y-\hat{H}x\|_{2}^{2}$, where $\hat{H}$ is an approximation based on partial knowledge of the matrix $H$. 
Also, we assume the decoder is unaware of the mismatch effect, and therefore, we set $\hat{H}=I$. 
The signal-to-noise ratio is defined as $\text{SNR} \triangleq 1/(2\sigma_{Z}^{2})$, and will be used in the subsequent simulations. 
\begin{figure}[ht]
	\centerline{\includegraphics[width=0.9\linewidth]{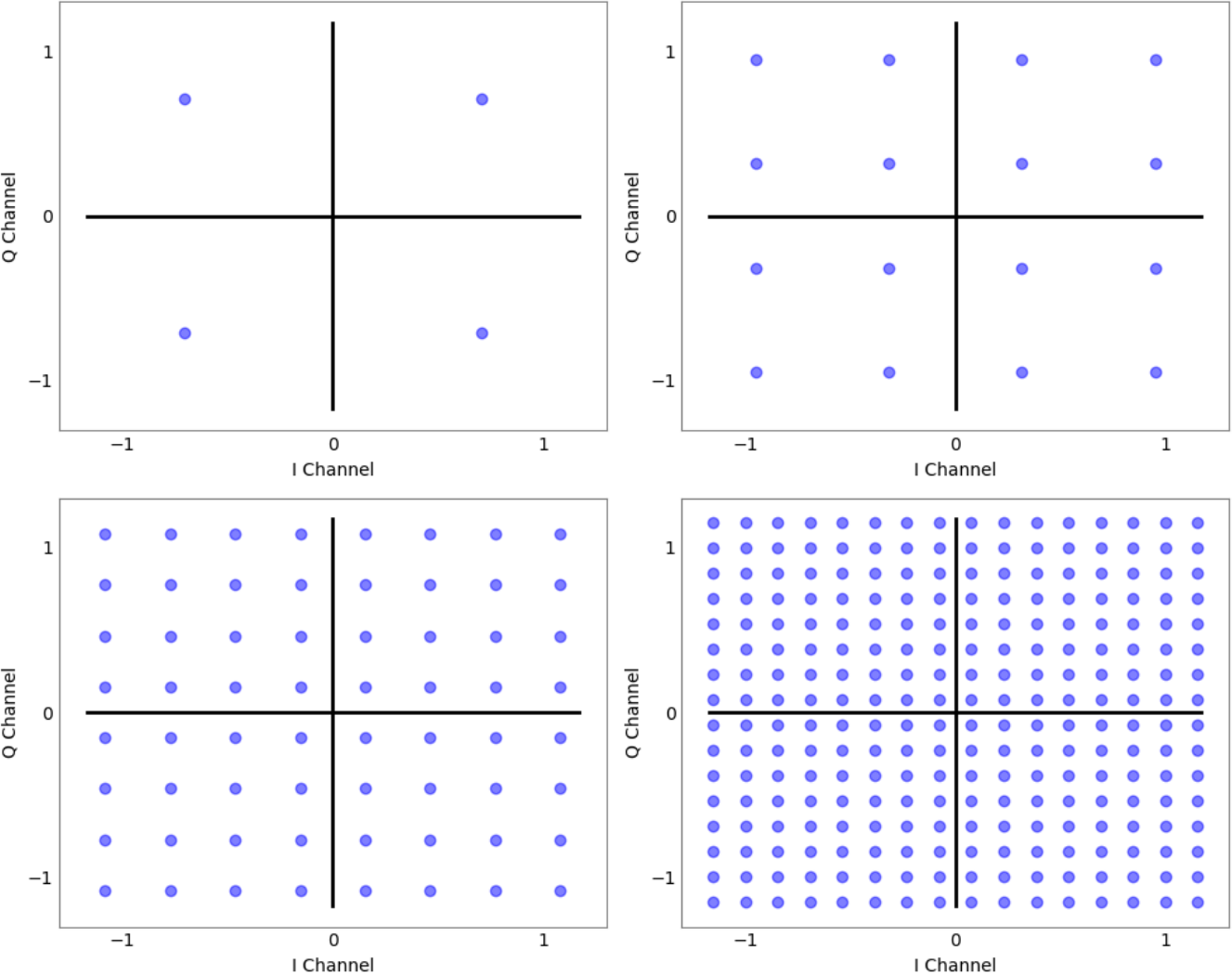}}
    \caption{Constellation points under the power constraint. 
    Upper Left: The QPSK modulation scheme. Upper Right: The 16-QAM modulation scheme. Lower Left: The 64-QAM modulation scheme. Lower Right: The 256-QAM modulation scheme.
    }
\label{Conspoint}
\end{figure}

For the simulations, four classical modulation schemes are taken into consideration, \thatis, QPSK, 16-QAM, 64-QAM, and 256-QAM. The constellation points (also referred to as the alphabet) for these schemes under the power constraint \eqref{pc} are illustrated in Fig. \ref{Conspoint}.

All the experiments are conducted on a platform with 128G RAM, and one Intel(R) Xeon(R) Gold 5117 CPU @2.00GHz with 14 cores.
%

\subsection{Algorithm Verification}
Specifically, for the additive white Gaussian noise (AWGN) channel, the parameter $\lambda$ in the Lagrangian \eqref{Lagrangian} can be updated by solving the equation \eqref{G_lambda}. 
This approach avoids the step-size tuning in the gradient projection step in line 8 of Algorithm \ref{alg:OT}, thereby augmenting the stability and robustness of our proposed algorithm. 
The validity of this iterative procedure is guaranteed by the proposition below.
\begin{proposition}\label{prop:g0}
    For the AWGN channel, 
    the equation $F(\lambda; \bm\phi, \bm\psi)=0$ with respect to $\lambda$ admits a unique solution on the interval $[0, +\infty)$. 
\end{proposition}

The proof of Proposition \ref{prop:g0} is provided in Appendix \ref{app:g0}.

\begin{figure}[ht]
	\centerline{\includegraphics[width=0.9\linewidth]{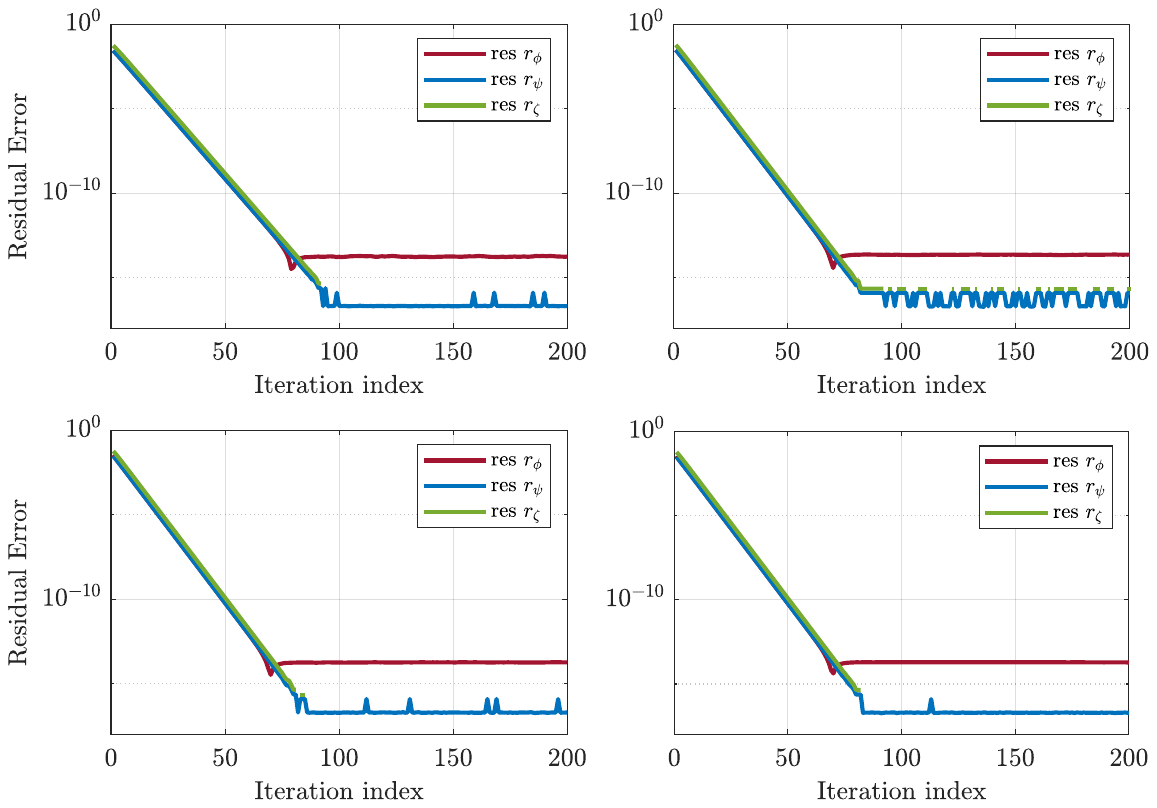}}
    \caption{The convergent trajectories of the residual error for $r_{\phi}$ (Red), $r_{\psi}$ (black) and $r_{\lambda}$ (Green). 
    Upper Left: The QPSK modulation scheme. Upper Right: The 16-QAM modulation scheme. Lower Left: The 64-QAM modulation scheme. Lower Right: The 256-QAM modulation scheme.}
    \label{Res}
\end{figure}

Accordingly, the convergence behavior of the proposed Sinkhorn-type algorithm is examined by evaluating the residual errors associated with updating equations \eqref{sink_iter} and \eqref{G_lambda}
\begin{equation*}
\begin{aligned}
r_{\phi} &= \sum_{i=1}^{M}|\phi_{i}\sum_{j=1}^{N}\Lambda_{ij}\psi_{j} - P_X(x_i)|,\\
r_{\psi} &= \sum_{j=1}^{N}|\psi_{j}\sum_{i=1}^{M}\Lambda_{ij}\phi_{i} - P_Y(y_j)|,\\
r_{\lambda} &= |F(\lambda; \bm\phi, \bm\psi)|.
\end{aligned}
\end{equation*}
The parameters are set to
\begin{equation} \label{expe01}
    (\eta,\theta) = (0.9, \pi/18), \; N = 250,000, \;
    \text{SNR} = 0 \textrm{dB}.
\end{equation}
Fig. \ref{Res} resents the convergence trajectories of the residual errors with respect to iteration steps. All three curves exhibit rapid decrease and reach machine-level precision within approximately $100$ iterations.
Notably, the proposed Sinkhorn-type algorithm demonstrates high computational efficiency.
In contrast to the classical optimal transport computation, which typically relies on entropy regularization approximations, the computation of \eqref{LM} inherently incorporates an entropy regularization term.
This structural property highlights the substantial efficiency advantage of our Sinkhorn-type algorithm.

\begin{table*}[ht]\label{compare}
    \renewcommand\arraystretch{1.25}
    \centering
    \caption{Comparison between the Sinkhorn-type algorithm and CVX. Columns 3-5 are the averaged computational time and the speed-up ratio of the Sinkhorn-type algorithm. Column 6 is the averaged difference of the LM rate computed by two methods.} 
    \label{Table_cvx} 
    \setlength{\tabcolsep}{6mm}{
    \begin{tabular}{c|c|c|c|c|c} 
        \toprule 
        \multirow{2}{*}{} & \multirow{2}{*}{$N$} & \multicolumn{2}{c|}{Computational time (s)} &
        \multirow{2}{*}{Speed-up ratio} & \multirow{2}{*}{Average difference}\\
        \cline{3-4}
         &  & Sinkhorn-type & CVX & &\\
        \hline  
        \multirow{3}{*}{QPSK} 
        & $100$ & $0.37\times10^{0}$ & $2.61\times10^{1}$ & $7.05\times10^{1}$ & $1.99\times10^{-7}$\\
        & $225$ & $0.64\times10^{0}$ & $8.66\times10^{1}$ & $1.35\times10^{2}$ & $6.45\times10^{-7}$\\
        & $400$ & $1.28\times10^{0}$ & - & - & -\\
        \hline
        \multirow{3}{*}{16-QAM} 
        & $100$ & $0.96\times10^{0}$ & $1.49\times10^{2}$ & $1.55\times10^{2}$ & $4.13\times10^{-7}$\\
        & $225$ & $1.96\times10^{0}$ & $9.27\times10^{2}$ & $4.73\times10^{2}$ & $1.37\times10^{-6}$\\
        & $400$ & $3.81\times10^{0}$ & - & - & -\\
        \hline
        \multirow{3}{*}{64-QAM} 
        & $100$ & $2.83\times10^{0}$ & $1.42\times10^{3}$ & $5.02\times10^{2}$ & $5.47\times10^{-7}$\\
        & $225$ & $6.09\times10^{0}$ & - & - & -\\ 
        & $400$ & $1.04\times10^{1}$ & - & - & -\\
        \hline
        \multirow{3}{*}{256-QAM} 
        & $100$ & $1.49\times10^{1}$ & - & - & -\\
        & $225$ & $2.32\times10^{1}$ & - & - & -\\
        & $400$ & $4.95\times10^{1}$ & - & - & -\\
        \bottomrule
        \multicolumn{6}{}{}\\[1pt]
    \end{tabular}}
\end{table*}
To further demonstrate the accuracy and efficiency of the proposed Sinkhorn-type algorithm for solving the LM rate problem \eqref{OT}, we compare it with the convex optimization toolbox, CVX \cite{2010cvx}, which serves as a baseline.
Table~\ref{Table_cvx} reports the average computational time and the average difference in optimal values obtained by two methods.
To mitigate the impact of randomness, each experiment is repeated with 100 times.
All parameters are consistent with those in \eqref{expe01}, except for $N$, which is varied to assess performance under different problem sizes.
Due to limitations of CVX in handling large-scale problems, we restrict $N$ to relatively small values, specifically $100$, $225$, and $400$.
The results indicate that both methods yield nearly identical optimal values.
However, the Sinkhorn-type algorithm significantly outperforms CVX in terms of computational speed, often by one or two orders of magnitude.
Furthermore, CVX fails to produce convergent results for moderately larger problem sizes, highlighting the scalability advantage of our approach.

\subsection{Results and Discussions}
The computational results of the LM rate are presented under various modulation schemes, channel parameters $(\eta,\theta)$, and signal-to-noise ratios (SNRs).
For comparison, the corresponding generalized mutual information (GMI) values \cite{zhang2011general} under identical configurations are also included. 
%
It is well known that GMI generally provides a lower bound on the LM rate \cite{2020Information}. 
%
These numerical results not only quantify the performance gap between LM rate and GMI but also serve to validate the accuracy and reliability of the proposed model and algorithm.

\begin{figure}[ht]
    \centerline{\includegraphics[width=0.9\linewidth]{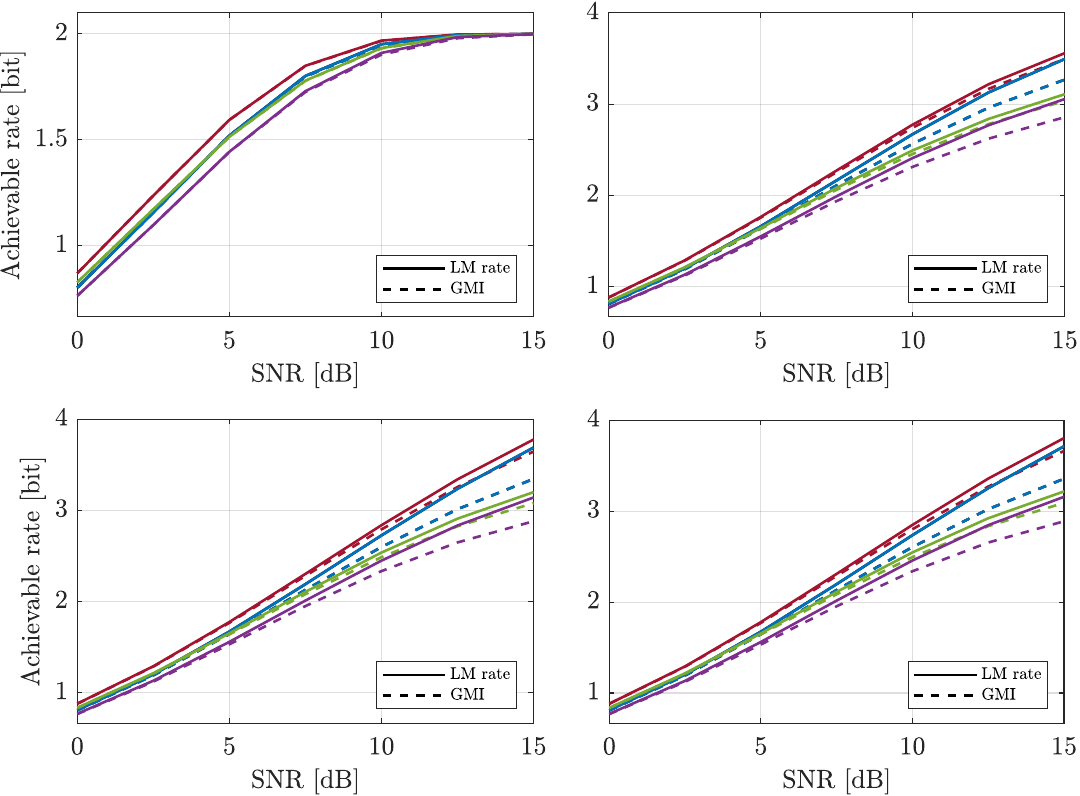}}
    \caption{LM rate (solid) and GMI (dashed) versus SNR  under different mismatched cases, including $(\eta,\theta) = (0.9, \pi/18)$ (Red), $(\eta,\theta) = (0.8, \pi/18)$ (black), $(\eta,\theta) = (0.9, \pi/12)$ (Green), and $(\eta,\theta) = (0.8, \pi/12)$ (Purple).
    Upper Left: The QPSK modulation scheme. Upper Right: The 16-QAM modulation scheme. Lower Left: The 64-QAM modulation scheme. Lower Right: The 256-QAM modulation scheme.}
    \label{QPAM}
\end{figure}

Four representative parameter settings are considered: \begin{equation*} (0.9, \frac{\pi}{18}),\quad (0.9, \frac{\pi}{12}),\quad (0.8, \frac{\pi}{18}), \quad (0.8, \frac{\pi}{12}). \end{equation*} To ensure discretization accuracy, $N$ is set to $250{,}000$, and each experiment is run for 500 iterations to guarantee the convergence of the algorithm.

Figure~\ref{QPAM} displays the LM rate and GMI as functions of SNR across different modulation schemes. The LM rate consistently exceeds GMI under all tested scenarios. In particular, the performance gain becomes more pronounced for $\text{SNR} > 8$ dB. Furthermore, both LM rate and GMI exhibit a decreasing trend as $\eta$ decreases (from $0.9$ to $0.8$) or $\theta$ increases (from $\pi/18$ to $\pi/12$), which aligns with theoretical expectations.

\section{Conclusion} \label{sec:conclusion}
In this paper, we studied the numerical computation of the LM rate, which is a lower bound for mismatch capacity. Our contributions are threefold. 
First, the LM rate was reformulated as an entropy-regularized optimal transport (OT) problem with an additional capacity constraint.
Second, we proposed a Sinkhorn-type algorithm tailored to this formulation. The algorithm consists of two key components: one part solves the OT structure efficiently through Sinkhorn iterations, while the other handles the extra constraint by reducing it to a
gradient projection step.
Third, an analytical framework was introduced to establish the convergence rate for the proposed algorithm, despite the challenges arising from the complex problem structure.
Numerical experiments demonstrate that the proposed method achieves high accuracy with significant computational efficiency. In addition, notable gains in the LM rate over the generalized mutual information (GMI) are observed under various modulation schemes and channel conditions.
These results suggest that the proposed OT-based approach offers promising potential for broader applications in mismatched decoding, as well as in the numerical computation of other bounds in information theory.


\begin{appendix}
\section{Proofs}

\subsection{Proof of Proposition \ref{prop:equi}}\label{app:equi}
\begin{proofs}
As it shown in \cite[p. 19]{2020Information}, the dual form of the LM rate can be expressed by the maximization problem \eqref{Scarlett_dual_LM_1}.
Using the notations adopted in this work, the expression \eqref{Scarlett_dual_LM_1} can be rewritten as:
\begin{equation} \label{Scarlett_dual_LM_2}
    \max_{\lambda,\widehat{\bdphi}}~~ \sum_{i=1}^{M}\sum_{j=1}^{N} P_X(x_i) W(y_j|x_i) \log\frac{e^{-\lambda d_{ij}} \widehat{\phi_{i}}}{\sum_{k=1}^{M} P_X(x_k) e^{-\lambda d_{kj}}\widehat{\phi_{k}}},
\end{equation}
where $\widehat{\phi_{k}} = e^{a(x_{k})}$ for simplicity.

On the other hand, substituting the dual variables $\bda, \bdb$ with $\bdphi, \bdpsi$ in our proposed dual form \eqref{dual_problem} yields the formulation
\begin{equation}\label{dual_new}
\begin{aligned}
    I_{\mathrm{LM}}(X, Y) =  \max_{\substack{\bm\phi, \bm\psi >\bm{0}, \\ \lambda > 0}}   ~
    \left(
    - \bm\phi^{T} \bm\Lambda \bm\psi + H(X) + H(Y) 
    \right. & \\
    \left.
    \mathbb{E}_{P_X}(\log \bm\phi) +  \mathbb{E}_{{P}_{Y}}(\log \bm\psi)   
    + \mathbb{E}_{P_{XY}}(\log \bm\Lambda) + 1
    \right) & ,
\end{aligned}
\end{equation}
where
$\bm\phi, \bm\psi, \Lambda$ are defined according to \eqref{dual_var}, and $H(\cdot)$ denotes the information entropy.
Actually, \eqref{dual_new} is the same as the \eqref{dual_problem}.
The function $g_{\mathrm{LM}}(\bdphi,\bdpsi,\lambda)$ is the objective function of our proposed dual form of LM rate \eqref{dual_new}

\vspace{-.1 in}
\begin{small}
\begin{align*}
    g_{\mathrm{LM}} (\bdphi ,\bdpsi ,\lambda) = 
    -\sum_{i=1}^M \sum_{j=1}^N \phi_i e^{-\lambda d_{ij}} \psi_j 
    +\sum_{i=1}^M P_X(x_i) \log \frac{\phi_i}{P_X(x_i)} & \\
    + \sum_{j=1}^N P_Y(y_j) \log \frac{\psi_j}{P_Y(y_j)} 
    - \lambda \sum_{i=1}^M \sum_{j=1}^N d_{ij}W(y_j|x_i)P_X(x_i) + 1
    & .
\end{align*}
\end{small}

Taking the derivative of $g_{\mathrm{LM}}(\bdphi,\bdpsi,\lambda)$ with respect to $\bdpsi$ leads to the condition
$\psi_{j}^{*} = {P_Y(y_j)}/{\left(\sum_{k=1}^{M} e^{-\lambda d_{kj}}\phi_{k}\right)}.$
%
Then, substituting this condition into $g_{\mathrm{LM}}(\bdphi,\bdpsi,\lambda)$, 
and denoting $\widehat{\phi_{i}} = {\phi_{i}}/{P_X(x_i)}$ further, we have 

\vspace{-.1in}
\begin{small}
\begin{equation*} \label{dual_LM_e}
\begin{aligned}
    g_{\mathrm{LM}}(\bdphi,\bdpsi^{*}\!,\lambda) 
    \!=\! \sum_{i=1}^{M}\! P_X(x_i) \log \widehat{\phi_i} \!-\! \lambda\sum_{i=1}^M \sum_{j=1}^N\!  d_{ij}W(y_j|x_i)P_X(x_i) & \\
    - \sum_{j=1}^N \left(\sum_{i=1}^M P_X(x_i) W(y_j|x_i)\right) \log\left(\sum_{k=1}^{M} P_X(x_k) e^{-\lambda d_{kj}} \widehat{\phi_{k}}\right) & .
\end{aligned}  
\end{equation*}
\end{small}

This is exactly the same as the objective in \eqref{Scarlett_dual_LM_2} by summing three terms and extracting the coefficients \!$P_X(x_i)W(y_j|x_i)$.
\end{proofs}

\subsection{Proof of Lemma \ref{lem:kernel}}\label{app:kernel}
\begin{proofs}
The kernel of matrix $\left(\bm K, \widetilde{\bm D}\right) \in \mbbR^{MN \times (M+N+1)}$ is defined as the notation, $\operatorname{ker}\left(\bm K, \widetilde{\bm D}\right)$, \thatis, 
\begin{equation*}
\begin{aligned}
    \left\{
    \left(\bm a; \bm b; c\right) \in \mathbb{R}^{(M+N+1)\times 1}  \mid 
    a_i+b_j+d_{ij}c=0, ~ \forall i,j \right\}.
\end{aligned}
\end{equation*}

When $z = 0$, the general solution of the above function $a_i+b_j+d_{ij}c = 0$ is clearly $a_i = -b_j = s$,
which yields
$$\operatorname{ker}(\bm K) = \{s\cdot(\mathbf{1}_{M}; -\mathbf{1}_{N})\in\mathbb{R}^{(M+N)\times 1} \mid s\in\mbbR\}.$$
Thus, we have $r(\bm K) = M + N-1$,
further leading to $r(\bm K, \widetilde{\bm D}) \ge M +N$. 

On the other hand, if we assume that the coefficient matrix $(\bm K, \widetilde{ \bm D})$ to be full rank, it derives 
$\bm K (\mathbf{p}; \mathbf{q}) = \widetilde{ \bm D}$, where $\mathbf{p}\in\mbbR^{M \times 1}, \mathbf{q}\in\mbbR^{N \times 1}$.
However, $\widetilde{ \bm D}$ is generated by the decoding metric. 
With the central symmetry of the alphabets in Assumption \ref{assp:03}, we could find a pair of $(m, n)$ satisfying
$$d_{11} = d_{mn}, ~ d_{1n} = d_{m1},~ d_{11} \neq d_{1n}.$$

Combining the above equations, we obtain
\begin{equation*}
    2d_{11} = p_1 + q_1 + p_m + q_n =  2d_{1n}, 
\end{equation*}
leading to a contradiction.
This suggests $r(\bm K, \widetilde{ \bm D}) \le M+N$.

In summary, we conclude that $r(\bm K, \widetilde{ \bm D}) = M+N,$
%
which means that the space $\operatorname{ker}(\bm K, \widetilde{ \bm D})$ is one-dimensional.
Note that $\operatorname{ker}(\bm K, \widetilde{ \bm D})$ contains an one-dimensional subspace $\{s\cdot(\mathbf{1}_{M}, \!-\mathbf{1}_{N}, 0), s\in\mbbR\}$, consequently leading to  \eqref{kernel}.  
\end{proofs}

\subsection{Proof of Lemma \ref{lem:lips}}\label{app:lips}
\begin{proofs}
%
%
Consider the function $h_\ell(\lambda) \triangleq g(\bda^{\ell+1}, \bdb^{\ell+1}, \lambda)$ at fixed dual variables $\bda^{\ell+1}, \bdb^{\ell+1}$ obtained from $(\ell+1)-$th iteration.
Accordingly, the second derivative of $h_\ell$ is
\begin{equation}\label{hessian}
\nabla_{\lambda}^2 h_\ell(\lambda) = \sum_{i=1}^M \sum_{j=1}^N d_{ij}^2 \exp\left(-\lambda d_{ij} \!-\! \alpha_i^{\ell+1} \!-\! \beta_j^{\ell+1} \!-\! 1\right).
\end{equation}

Denote $Q_{ij}^{\ell+1/2} \triangleq \mrme^{-\lambda^\ell d_{ij} - \alpha_i^{\ell+1} - \beta_j^{\ell+1} - 1}$, which satisfies
\begin{equation}\label{eq:finite}
    \sum_{i=1}^M \sum_{j = 1}^N Q_{ij}^{\ell+1/2} = \sum_{i=1}^M \sum_{j = 1}^N\mrme^{-\lambda^\ell d_{ij} - \alpha_i^{\ell+1} - \beta_j^{\ell+1} - 1} = 1,
\end{equation}
by the optimality of $\bda^{\ell+1}, \bdb^{\ell+1}$. 
Also, let $\Delta_\ell \triangleq \lambda^{\ell+1} - \lambda^\ell$ and $M_{D} \triangleq \|\bm D\|_{\infty}$ where $\bm D = \left(d_{ij}\right)_{M\times N}$ is the decoding metric.
For any $\lambda$ in the interval $I_\ell = [\min(\lambda^\ell, \lambda^{\ell+1}), \max(\lambda^\ell, \lambda^{\ell+1})]$, 
%
\begin{align*}
\exp(-\lambda d_{ij}) 
&= \exp\left(-\lambda^\ell d_{ij} + (\lambda^\ell - \lambda)d_{ij}\right) \nonumber \\
&\leq \exp\left(-\lambda^\ell d_{ij}\right) \exp\left(|\lambda^\ell - \lambda| \cdot |d_{ij}|\right) \nonumber \\
&\leq \exp\left(-\lambda^\ell d_{ij}\right) \exp\left(|\lambda^\ell - \lambda^{\ell+1}| \cdot M_{D}\right).
\end{align*}
%

Substituting it into \eqref{hessian} yields
\begin{equation}
\begin{aligned}
    \nabla_{\lambda}^2 h_\ell(\lambda) 
    &\leq \sum_{i=1}^M \sum_{j = 1}^N d_{ij}^2 Q_{ij}^{\ell+1/2} \exp\left(|\lambda^\ell - \lambda^{\ell+1}| \cdot M_{D}\right)  \\
    &\leq M_{D}^2 \exp\left(|\lambda^\ell - \lambda^{\ell+1}| \cdot M_{D}\right) \sum_{i=1}^M \sum_{j = 1}^N Q_{ij}^{\ell+1/2} \\
    &= M_{D}^2 \exp\left(|\Delta_\ell| M_{D}\right).
\end{aligned}
\end{equation}

Accordingly, $M_{D}^2 \exp\left(|\Delta_\ell| M_{D}\right)$ provides a uniform upper bound that is independent of the specific values of $\bda^{\ell+1}$, $\bdb^{\ell+1}$ at $(\ell+1)-$th iteration.
Then, for any $\lambda_1, \lambda_2 \in I_\ell$, we have
\begin{align*}
|\nabla_\lambda h_\ell(\lambda_1) - \nabla_\lambda h_\ell(\lambda_2)| 
&\leq \sup_{\xi \in I_\ell} |\nabla_\lambda^2 h_\ell(\xi)| \cdot |\lambda_1 - \lambda_2| \nonumber \\
&\leq M_{D}^2 \exp(|\Delta_\ell| M_{D}) |\lambda_1 - \lambda_2|.
\end{align*}
Thus, $h_\ell(\lambda)$ is $L_{\lambda}^{\ell}$-smooth on $I_\ell$ with a local Lipschitz constant
\begin{equation}
L_{\lambda}^{\ell} \triangleq M_{D}^2 \exp(|\Delta_\ell| M_{D}).
\end{equation}

The update $\lambda^{\ell+1} = [\lambda^\ell - \tau \nabla_\lambda h_\ell(\lambda^\ell)]_+$ ensures $\tau$ is chosen such that $\lambda^{\ell+1}$ remains in a region where the gradient change is controlled. 
%
Also, the gradient during iteration 
\begin{equation*}
\begin{aligned}
    & \nabla_\lambda h_\ell(\lambda^\ell) 
    =
    \nabla_{\lambda}^2g(\bda^{\ell+1},\bdb^{\ell+1},\lambda^{\ell}) 
    \\
    & \quad
     = \! \sum_{i=1}^M \! \sum_{j=1}^N \! d_{ij} \exp{\left(-\lambda^{\ell} d_{ij}-\alpha_i^{\ell+1}-\beta_j^{\ell+1}-1\right)} \\
     & \quad \leq M_D \sum_{i=1}^M \sum_{j = 1}^N Q_{ij}^{\ell+1/2} = M_D
\end{aligned}
\end{equation*}
is finite due to the derived condition \eqref{eq:finite}.
For sufficiently small $\tau$ (e.g., $\tau \leq 1/M_{D}^2$), we have $|\Delta_\ell| = O(\tau)$, which yields an upper bound $\Delta = \sup_l \vert\Delta_\ell\vert$.
Then 
$$ L_{\lambda} \triangleq \sup_{\ell} L_{\lambda}^{\ell} \leq M_{D}^2 \exp(\Delta M_{D})$$
remains bounded. 
Hence,
\begin{equation*}
\begin{aligned}
    \|\nabla h_{\ell}(\lambda_{1}) -\nabla h_{\ell}(\lambda_{2})\|_{2}
    &\leq L_{\lambda}\|\lambda_1 - \lambda_2\|_{2}.
\end{aligned}
\end{equation*}
The desired inequality then follows,
with $L_{\lambda}$ uniformly bounded under the proper step size selection.
\end{proofs}

\subsection{Proof of Lemma \ref{lem:grad}}\label{app:grad}
\begin{proofs}
According to the updating rule of $\lambda^{\ell+1}$, the property of the projection operator yields
\begin{equation}\label{eq65:lam}
    \left \langle \lambda^{\ell} - \tau \nabla_{\lambda} g(\bda^{\ell+1}, \bdb^{\ell+1}, \lambda^{\ell}) - \lambda^{\ell+1}, z - \lambda^{\ell+1} \right \rangle \leq 0,
\end{equation}
for any $z\in [0. \infty)$. Substituting $z = \lambda^{\ell}$ in \eqref{eq65:lam}, we have
\begin{equation}\label{eq66:lam}
\begin{aligned}
    \left \langle \nabla_{\lambda} g(\bda^{\ell+1}, \bdb^{\ell+1}, \lambda^{\ell}), \lambda^{\ell} - \lambda^{\ell+1} \right \rangle 
    \geq \frac{1}{\tau} \| \lambda^{\ell} - \lambda^{\ell+1} \|_{2}^2. 
\end{aligned} 
\end{equation}
By utilizing \eqref{L-smooth} in Lemma \ref{lem:lips}, and recalling $\tau \leq \frac{1}{L_{\lambda}}$, we obtain
\begin{equation*}
\begin{aligned}
    & g(\bda^{\ell+1}, \bdb^{\ell+1}, \lambda^{\ell+1}) \leq g(\bda^{\ell+1}, \bdb^{\ell+1}, \lambda^{\ell})  \\
    &\quad~ + \langle \nabla_{\lambda}g(\bda^{\ell+1}, \bdb^{\ell+1}, \lambda^{\ell}), \lambda^{\ell+1}-\lambda^{\ell} \rangle+\frac{L_{\lambda}}{2}\|\lambda^{\ell+1}-\lambda^{\ell}\|^{2}_{2} \\
    &\quad \leq  g(\bda^{\ell+1}, \bdb^{\ell+1}, \lambda^{\ell}) - \frac{L_{\lambda}}{2}\|\lambda^{\ell+1}-\lambda^{\ell}\|^{2}_{2}.
\end{aligned}
\end{equation*}
Then the above inequality admits 
the desired result, \thatis
\begin{equation*}\label{eq67:lam}
    g(\bda^{\ell+1}, \bdb^{\ell+1}, \lambda^{\ell}) - g(\bda^{\ell+1}, \bdb^{\ell+1}, \lambda^{\ell+1}) \geq \frac{L_{\lambda}}{2}\|\lambda^{\ell+1}-\lambda^{\ell}\|^{2}_{2}.
\end{equation*}
%

Note that the objective function $g$ strictly decreases at each step of the alternating iteration \eqref{update_notation}.
This ensures the sequence of iterates $(\bda^{\ell},\bdb^{\ell},\lambda^{\ell})$ converges.
Critically, the optimal multiplier $\lambda^{*}$ is unique and finite. 
Hence, the sequence $\{\lambda^{\ell}\}$ is bounded and converges to $\lambda^{*}$.

Accordingly, there exists a finite constant $M_{\lambda}\triangleq\sup_{\ell}\lambda^{\ell}/2$. 
Then we propose the estimation of $\lambda$ apparently by
\begin{equation*}
    \sup_{\ell} \vert\lambda^{\ell}-\lambda^{*}\vert \leq M_{\lambda}.
\end{equation*}


For the gradient inequality of $\lambda$, the left hand side is split into three parts
\begin{equation}\label{eq:gplit}
\begin{aligned}
    &\left \langle \nabla_{\lambda} g(\bda^{\ell+1}, \bdb^{\ell}, \lambda^{\ell}), \lambda^{\ell} - \lambda^{*} \right \rangle \\
    &\quad= 
    \left \langle \nabla_{\lambda} g(\bda^{\ell+1}, \bdb^{\ell}, \lambda^{\ell}) - \nabla_{\lambda} g(\bda^{\ell+1}, \bdb^{\ell+1}, \lambda^{\ell}), \lambda^{\ell} - \lambda^{*} \right \rangle \\
    &\quad\quad+ \left \langle \nabla_{\lambda} g(\bda^{\ell+1}, \bdb^{\ell+1}, \lambda^{\ell}), \lambda^{\ell} - \lambda^{\ell+1} \right \rangle \\
    &\quad\quad+ \left \langle \nabla_{\lambda} g(\bda^{\ell+1}, \bdb^{\ell+1}, \lambda^{\ell}), \lambda^{\ell+1} - \lambda^{*} \right \rangle.
\end{aligned}
\end{equation}

For the first term in \eqref{eq:gplit},
\begin{equation}
\begin{aligned}
    & \|\nabla_{\lambda} g(\bda^{\ell+1}, \bdb^{\ell}, \lambda^{\ell}) - \nabla_{\lambda} g(\bda^{\ell+1}, \bdb^{\ell+1}, \lambda^{\ell})\|_{1} \\
    & = \|\mathbf{1}_{M}^{\top} \mathcal{Q}(\bda^{\ell+1}, \bdb^{\ell}, \lambda^{\ell}) \mathbf{1}_{N} 
    - \mathbf{1}_{M}^{\top} \mathcal{Q}(\bda^{\ell+1}, \bdb^{\ell+1}, \lambda^{\ell}) \mathbf{1}_{N} \|_1 \\
    &  = \Big\|\left \langle \mathbf{1}_{N} - \mrme^{\bdb^{\ell}-\bdb^{\ell+1}}, \mathcal{Q}(\bda^{\ell+1}, \bdb^{\ell}, \lambda^{\ell})^{\top}\mathbf{1}_{M} \right \rangle\Big\|_{1} 
    \\
    &  = \!\Big\|\left \langle \mathbf{1}_{N} \!-\! \frac{P_{Y}(\bdy)}{\mathcal{Q}(\bda^{\ell+1}, \bdb^{\ell}, \lambda^{\ell})^{\top}\mathbf{1}_{M}}, \mathcal{Q}(\bda^{\ell+1}, \bdb^{\ell}, \lambda^{\ell})^{\top}\mathbf{1}_{M} \right \rangle\Big\|_{1} \\
    &  = \|\nabla_{\bdb} g(\bda^{\ell+1}, \bdb^{\ell}, \lambda^{\ell})\|_{1}.
\end{aligned}
\end{equation}
where the third equation is based on \eqref{updating_ab}. Further, 
\begin{equation}\label{eq:gp1}
\begin{aligned}
    & \left \langle \nabla_{\lambda} g(\bda^{\ell+1}, \bdb^{\ell}, \lambda^{\ell}) - \nabla_{\lambda} g(\bda^{\ell+1}, \bdb^{\ell+1}, \lambda^{\ell}), \lambda^{\ell} - \lambda^{*} \right \rangle \\
    & \quad \leq \|\nabla_{\bdb} g(\bda^{\ell+1}, \bdb^{\ell}, \lambda^{\ell})\|_{1} \|\lambda^{\ell} - \lambda^{*}\|_{\infty}  \\
    & \quad \leq  M_{\lambda}\|\nabla_{\bdb} g(\bda^{\ell+1}, \bdb^{\ell}, \lambda^{\ell})\|_{1}.
\end{aligned}
\end{equation}

For the second term in \eqref{eq:gplit}, recalling $T = \mathbb{E}_{P_{XY}}(d(X,Y))$ 
and $\mathbf{1}_{M}^{\top} \mathcal{Q}(\bda^{\ell+1}, \bdb^{\ell+1}, \lambda^{\ell}) \mathbf{1}_{N}$, we have
\begin{equation*}
\begin{aligned}
    &\|\nabla_{\lambda} g(\bda^{\ell+1}, \bdb^{\ell+1}, \lambda^{\ell})\|_{2} 
    = \vert T - \left \langle \mathcal{Q}(\bda^{\ell+1}, \bdb^{\ell+1}, \lambda^{\ell}), \bm D \right\rangle \vert \\
    &\leq \!\mathbb{E}_{P_{XY}}\!(d(X,Y)) \!+\! \|\mathcal{Q}(\bda^{\ell+1},\!\bdb^{\ell+1}, \!\lambda^{\ell})\|_1 \|\bm D\|_{\infty} 
    \!\leq\! 2 \|\bm D\|_{\infty}.
\end{aligned}
\end{equation*}
%
%
Using Cauchy-Schwartz inequality, we have
\begin{equation}\label{eq:gp2}
\begin{aligned}
    & \left \langle \nabla_{\lambda} g(\bda^{\ell+1}, \bdb^{\ell+1}, \lambda^{\ell}), \lambda^{\ell} - \lambda^{\ell+1} \right \rangle \\
    &\quad\quad \leq \|\nabla_{\lambda} g(\bda^{\ell+1}, \bdb^{\ell+1}, \lambda^{\ell})\|_{2} \|\lambda^{\ell} - \lambda^{\ell+1}\|_{2} \\
    &\quad\quad \leq 2 M_{D}  \|\lambda^{\ell} - \lambda^{\ell+1}\|_{2}.
\end{aligned}
\end{equation}

For the third term in \eqref{eq:gplit}, substituting $z = \lambda^{*}$ in \eqref{eq65:lam} yields 
\begin{equation}\label{eq2057:lam}
\begin{aligned}
    &\left \langle \nabla_{\lambda} g(\bda^{\ell+1}, \bdb^{\ell+1}, \lambda^{\ell}), \lambda^{\ell+1} - \lambda^{*} \right \rangle \\
    &\quad\quad \leq \frac{1}{\tau} \left \langle \lambda^{\ell} - \lambda^{\ell+1}, \lambda^{\ell+1} - \lambda^{*} \right \rangle 
    \leq \frac{M_{\lambda}}{\tau}  \|\lambda^{\ell} - \lambda^{\ell+1}\|_{2}. 
\end{aligned}
\end{equation}

Set $M_0 \triangleq \max \{\frac{M_{\lambda}}{\tau L_{\lambda}}, \frac{2 M_D}{L_{\lambda}} \}$ to simplify the coefficient in the bound.

As $\tau \leq \frac{1}{L_{\lambda}}$, this choice guarantees $M_{\lambda} \leq \frac{M_{\lambda}}{\tau L_{\lambda}} \leq M_{0}$.
Combining \eqref{eq:gplit}, \eqref{eq:gp1}, \eqref{eq:gp2} and \eqref{eq2057:lam}, the proof is complete.

\end{proofs}

\subsection{Proof of Lemma \ref{lem:bound}} \label{app:bound}
\begin{proofs}
%
%
We focus on the bounds for $\bda^{\ell},\bdb^{\ell}$. 
Denoting
$C_{d}\triangleq\min_{i,j}\mrme^{-2d_{ij} M_{\lambda}}$, $M_{\lambda}=\sup_{\ell} \lambda^{\ell}/2$,
the following inequalities on $P_X(x),P_Y(y)$ hold 
\begin{equation}\label{R_left}
\begin{aligned}
    & P_X(x_i)  
    = \sum_{j=1}^{N} \phi_{i}^{\ell+1}\mrme^{-\lambda^{\ell} d_{ij}}\psi_{j}^{\ell} 
    \\ & \quad\quad
    \geq \mrme^{-\alpha^{\ell+1}_{i} - 1/2} C_{d} \left\langle\mathbf{1}_{N}, \exp(-\bdb^{\ell}-\mathbf{1}_{N}/2) \right\rangle, \\
    & P_Y(y_j) 
    = \sum_{i=1}^{M} \psi_{j}^{\ell+1}\mrme^{-\lambda^{\ell} d_{ij}}\phi_{j}^{\ell+1}
    \\ & \quad\quad
    \geq \mrme^{-\beta^{\ell+1}_{j}-1/2} C_{d} \left\langle\mathbf{1}_{M}, \exp(-\bda^{\ell+1} \!-\! \mathbf{1}_{M}/2) \right\rangle.
\end{aligned}
\end{equation}
With the condition 
$0<P_X(x_i),P_Y(y_j)\leq 1$ in Assumption \ref{assp:02},
a further estimation is generated from \eqref{R_left} as follows
\begin{equation}\label{R_left_1}
\begin{aligned}
    \alpha_{i}^{\ell+1} & \geq -\ln\Big(\frac{\max_{i}P_X(x_i)}{C_{d}\left\langle\mathbf{1}_{N}, \exp(-\bdb^{\ell}-\mathbf{1}_{N}/2) \right\rangle}\Big) -\frac{1}{2}, 
    \\
    \beta_{i}^{\ell+1} & \geq -\ln\Big(\frac{\max_{j}P_Y(y_j)}{C_{d}\left\langle\mathbf{1}_{M}, \exp(-\bda^{\ell+1}-\mathbf{1}_{M}/2) \right\rangle}\Big) -\frac{1}{2}.
\end{aligned}
\end{equation}

On the other hand, with $\mrme^{-\lambda d_{ij}}\leq 1$, we can obtain the reverse inequalities
\begin{equation*}\label{R_right}
\begin{aligned}
    & P_X(x_i) 
    \leq \mrme^{-\alpha^{\ell+1}_{i} - 1/2} \left\langle\mathbf{1}_{N}, \exp(-\bdb^{\ell}-\mathbf{1}_{N}/2) \right\rangle, \\
    & P_Y(y_j) 
    \leq \mrme^{-\beta^{\ell+1}_{j} - 1/2} \left\langle\mathbf{1}_{M}, \exp(-\bda^{\ell+1}-\mathbf{1}_{M}/2) \right\rangle. 
\end{aligned}
\end{equation*}
Similarly, combine the above estimation with 
$0<P_X(x_i)$, $P_Y(y_j)\leq 1$
, we further obtain 
\begin{equation}\label{R_right_1}
\begin{aligned}
    \alpha_{i}^{\ell+1} 
    & \leq -\ln\Big(\frac{\min_{i}P_X(x_i)}{\left\langle\mathbf{1}_{N}, \exp(-\bdb^{\ell}-\mathbf{1}_{N}/2) \right\rangle}\Big) -\frac{1}{2}, 
    \\ 
    \beta_{j}^{\ell+1} 
    & \leq -\ln\Big(\frac{\min_{j}P_Y(y_j)}{\left\langle\mathbf{1}_{M}, \exp(-\bda^{\ell+1}-\mathbf{1}_{M}/2) \right\rangle}\Big) -\frac{1}{2}.
\end{aligned}
\end{equation}
Therefore, sum up \eqref{R_left_1} and \eqref{R_right_1}, we have the finial estimation
\begin{equation*}
\begin{aligned}
    &\max_{i}\alpha_{i}^{\ell+1}-\min_{i}\alpha_{i}^{\ell+1} \leq -\ln\Big(C_{d}\frac{\min_{i}P_X(x_i)}{\max_{i}P_X(x_i)}\Big), \\
    &\max_{j}\beta_{j}^{\ell+1}-\min_{j}\beta_{j}^{\ell+1} \leq -\ln\Big(C_{d}\frac{\min_{j}P_Y(y_j)}{\max_{j}P_Y(y_j)}\Big).
\end{aligned}
\end{equation*}

Recalling that $\sup_{\ell} \vert\lambda^{\ell}-\lambda^{*}\vert \leq M_{\lambda} \leq M_0$,
the proof is finished by defining the upper bound $S_{0}$ as
\begin{equation}\label{R_definition}
\begin{aligned}
    & S_{0} = \max\Bigg\{ M_{0}, \\ 
    & ~ 
    -\ln\bigg(C_{d}\cdot\min\Big\{\frac{\min_{i}P_X(x_i)}{\max_{i}P_X(x_i)},\frac{\min_{j}P_Y(y_j)}{\max_{j}P_Y(y_j)}\Big\}\bigg)\Bigg\}.
\end{aligned}
\end{equation}
\end{proofs}

\subsection{Proof of Lemma \ref{lem:decr}}\label{app:decr}
\begin{proofs}
The updating rules of $\bda^{\ell+1}$ and $\bdb^{\ell+1}$ yield that 
\begin{equation}\label{eq:cddd}
    0 = \nabla_{\bda}g(\bda^{\ell+1}, \bdb^{\ell}, \lambda^{\ell}) = \nabla_{\bdb}g(\bda^{\ell+1}, \bdb^{\ell+1}, \lambda^{\ell}).
\end{equation}
Accordingly, we have
\begin{equation}\label{eq:sum1}
\begin{aligned}
    \mathbf{1}_{M}^{\top} \mathcal{Q}(\bda^{\ell+1}, \bdb^{\ell}, \lambda^{\ell}) \mathbf{1}_{N} &= \mathbf{1}_{M}^{\top}P_X(\bdx) = 1, \\
    \mathbf{1}_{M}^{\top} \mathcal{Q}(\bda^{\ell+1}, \bdb^{\ell+1}, \lambda^{\ell}) \mathbf{1}_{N} &= P_Y(\bdy)^{\top}\mathbf{1}_{N} = 1.
\end{aligned}
\end{equation}

Thus, the decrease on the direction of $\bdb$ is
\begin{equation*}
\begin{aligned}
    & g(\bda^{\ell+1}, \bdb^{\ell}, \lambda^{\ell}) - g(\bda^{\ell+1}, \bdb^{\ell+1}, \lambda^{\ell}) \\
    & \quad = 
    \mathbf{1}_{M}^{\top} \mathcal{Q}(\bda^{\ell+1}, \bdb^{\ell}, \lambda^{\ell}) \mathbf{1}_{N}
    - \mathbf{1}_{M}^{\top} \mathcal{Q}(\bda^{\ell+1}, \bdb^{\ell+1}, \lambda^{\ell}) \mathbf{1}_{N}
    \\
    & \quad\quad + \left\langle \bdb^{\ell} - \bdb^{\ell+1}, P_Y(\bdy) \right\rangle \\
    & \quad = \left\langle \ln P_Y(\bdy) - \ln (\mathcal{Q}(\bda^{\ell+1}, \bdb^{\ell}, \lambda^{\ell})^{\top} \mathbf{1}_{M}), P_Y(\bdy) \right\rangle \\
    & \quad \geq \frac{1}{2}\|(\mathcal{Q}(\bda^{\ell+1}, \bdb^{\ell}, \lambda^{\ell})^{\top}\mathbf{1}_{M}-P_Y(\bdy)\|_{1}^2 ,
\end{aligned}
\end{equation*}
where the last inequality holds due to the Pinsker inequality.
Hence we obtain the reduction inequality \eqref{esti_beta}.

Subsequently, we focus on the estimation of the gradient. Recalling the equation \eqref{eq:sum1}, we have
\begin{equation*}
    \left\langle(\mathcal{Q}(\bda^{\ell + 1}, \bdb^{\ell}, \lambda^{\ell}))^{\top}\mathbf{1}_{M}-P_Y(\bdy), \mathbf{1}_{N}\right\rangle = 0.
\end{equation*}
Then for any $\eta\in\mbbR$, the following inequality holds
\begin{equation}\label{tau_1}
\begin{aligned}
    & \left \langle \nabla_{\bdb} g(\bda^{\ell+1}, \bdb^{\ell}, \lambda^{\ell}), \bdb^{\ell}\right \rangle \\
    & \quad\quad = \left \langle P_Y(\bdy) - \mathcal{Q}(\bda^{\ell + 1}, \bdb^{\ell}, \lambda^{\ell}))^{\top}\mathbf{1}_{M}, \bdb^{\ell} -\eta\mathbf{1}_{N} \right \rangle \\
    & \quad\quad  \leq \|\nabla_{\bdb} g(\bda^{\ell+1}, \bdb^{\ell}, \lambda^{\ell})\|_{1} \|\bdb^{\ell}-\eta\mathbf{1}_{N}\|_{\infty}.
\end{aligned}
\end{equation}
Similar as the technique in \cite[Lemma 1]{2018Dvurechensky}, substituting $\eta={(\max_{i}\bdb^{\ell}_{i}+\min_{i}\bdb^{\ell}_{i})}/{2}$ in \eqref{tau_1} yields that 
\begin{equation}\label{tau_2}
    \|\bdb^{\ell}-\eta\mathbf{1}_{N}\|_{\infty} \leq \frac{\max_{i}\bdb^{\ell}_{i}-\min_{i}\bdb^{\ell}_{i}}{2} \leq \frac{S_{0}}{2},
\end{equation}
where $S_{0}$ is defined in Lemma \ref{lem:bound}. 
Then with the combination of \eqref{tau_1} and \eqref{tau_2}, we obtain 
\begin{equation}\label{tau_3}
    \left \langle \nabla_{\bdb} g(\bda^{\ell+1}, \bdb^{\ell}, \lambda^{\ell}), \bdb^{\ell}\right \rangle \leq \frac{S_{0}}{2} \|\nabla_{\bdb} g(\bda^{\ell+1}, \bdb^{\ell}, \lambda^{\ell})\|_{1}.
\end{equation}

Replacing $\bdb^{\ell}$ with $\bdb^{*}$ in the \eqref{tau_2}, we can obtain an inequality with the similar structure
\begin{equation}\label{tau_4}
\begin{aligned}
    & \left \langle - \nabla_{\bdb} g(\bda^{\ell+1}, \bdb^{\ell}, \lambda^{\ell}), \bdb^{*}\right \rangle 
    \\
    & \quad\quad
    \leq \|- \nabla_{\bdb} g(\bda^{\ell+1}, \bdb^{\ell}, \lambda^{\ell})\|_{1} \|\bdb^{*}-\eta\mathbf{1}_{N}\|_{\infty} \\
    & \quad\quad  
    \!\leq\! \frac{S_{0}}{2} \|\nabla_{\bdb} g(\bda^{\ell+1}, \bdb^{\ell}, \lambda^{\ell})\|_{1}.
\end{aligned}
\end{equation}
Therefore, with the combination of \eqref{tau_3}, and \eqref{tau_4}, the left hand side of the expected inequality \eqref{gddd_beta} is proposed as
\begin{equation*}
\begin{aligned}
    &\left \langle \nabla_{\bdb} g(\bda^{\ell+1}, \bdb^{\ell}, \lambda^{\ell}), \bdb^{\ell} - \bdb^{*} \right \rangle  \\
    & \quad = \left \langle \nabla_{\bdb} g(\bda^{\ell+1}, \bdb^{\ell}, \lambda^{\ell}), \bdb^{\ell}\right \rangle + \left \langle -\nabla_{\bdb} g(\bda^{\ell+1}, \bdb^{\ell}, \lambda^{\ell}), \bdb^{*}\right \rangle \\
    & \quad \leq S_{0} \|\nabla_{\bdb} g(\bda^{\ell+1}, \bdb^{\ell}, \lambda^{\ell+1})\|_{1}.
\end{aligned}
\end{equation*}
\end{proofs}

%

\subsection{Proof of Proposition \ref{prop:g0}}\label{app:g0}
\begin{proofs}
We first claim that the inequality constraint \eqref{LM_ineq} is equivalent to
\begin{equation}\label{6d:inner}
\int_{\mcX}\int_{\mcY} {Q}_{X Y}(x, y) \langle x,y\rangle \mrmd x\mrmd y 
\geq 
\int_{\mcX}P_{X}(x) \langle Hx,x\rangle \mrmd x.
\end{equation}

In Proposition \ref{prop:g0}, we focus on the AWGN channel described by the following condition
\begin{equation*}
    Y = H X + Z, \ \text{with } Z\sim \mathcal{N}(0,\sigma_{Z}^{2}).
\end{equation*}
Specially, the transition law is $W(y|x) = \frac{1}{2\pi \sigma_{Z}^{2}} e^{-\frac{\| y - Hx\|^2}{2\sigma_{Z}^{2}}}$,
and the decoding metric is $d(x,y) = \|y - x\|^2$.
Also, we denote $z = y - Hx \in \mathcal{Z}$ for simplicity.
Then the left hand side of the inequality constraint \eqref{LM_ineq} can be written as

\vspace{-.1 in}
\begin{small}
\begin{equation}\label{6d:LHS}
\begin{aligned}
    & \text{LHS} = \int_{\mcX}\int_{\mathcal{Z}} P_X(x) \frac{1}{2\pi \sigma_{Z}^{2}} e^{-\frac{\| z \|^2}{2\sigma_{Z}^{2}}} \| z\|^2 \mrmd x\mrmd z \\
    & +\!\!\int_{\mcX} P_X(x) \left( \|Hx\|^2 + \|x\|^2 \right) \mrmd x 
    \!-\! 2 \int_{\mcX} \! \int_{\mcY} {Q}_{X Y}(x, y) \langle x,y\rangle \mrmd x\mrmd y.
\end{aligned}
\end{equation}
\end{small}
\!\!\!\!
Meanwhile, the right hand side of the inequality constraint \eqref{LM_ineq} can be written as

\vspace{-.1in}
\begin{small}
\begin{equation}\label{6d:RHS}
\begin{aligned}
    T 
    = & \!\int_{\mcX} \! \int_{\mathcal{Z}} P_X(x) \frac{e^{-\frac{\| z \|^2}{2\sigma_{Z}^{2}}}}{2\pi \sigma_{Z}^{2}}  \| z\|^2 \mrmd x\mrmd z 
    \! +\! \int_{\mcX} P_X(x) \|Hx -x\|^2 \mrmd x.
\end{aligned}
\end{equation}
\end{small}
\!\!\!\!
The inequality \eqref{6d:inner} is obtained by combining the equality \eqref{6d:LHS}, \eqref{6d:RHS}, and the inequality \eqref{LM_ineq}.

Based on \eqref{6d:inner}, the condition $F(\lambda^{\ell}; \bm\phi^{\ell}, \bm\psi^{\ell}) \geq 0$ is equivalent to $\hat{F}^{\ell}(\lambda^{\ell}; \bm\phi^{\ell}, \bm\psi^{\ell}) \geq 0$ with
\begin{equation*}
\begin{aligned}
    \hat{F}^{\ell}
    \triangleq & 
    \sum_{i=1}^M P_{X}(x_i) \langle Hx_i, x_i \rangle 
    - \sum_{i=1}^{M}\sum_{j=1}^{N} \langle x_i, y_j \rangle \phi_{i}^{\ell}\psi_{j}^{\ell} \mrme^{\lambda \langle x_i, y_j \rangle},
\end{aligned}
\end{equation*}
where $\ell$ is the iteration number, and

\vspace{-.1in}
\begin{small}
\begin{equation*}
    \phi_i^{\ell} = \frac{P_X(x_{i})} {\sum_{j = 1}^N \psi_{j}^{\ell - 1} \mrme^{\lambda \langle x_{i}, y_{j} \rangle}}, ~
    \psi_j^{\ell} = \frac{P_Y(y_{j})}  {\sum_{i = 1}^M \phi_{i}^{\ell - 1} \mrme^{\lambda \langle x_{i}, y_{j} \rangle}}.
\end{equation*}
\end{small}

Following the central symmetry of alphabets in Assumption \ref{assp:03}, symmetric points of $x_i$ and $y_j$ with respect to the coordinate origin are denoted by $x_{M+1-i}$ and $y_{N+1-j}$, respectively. Then, we have
\begin{equation*}
     \langle x_{i}, y_{j} \rangle =  \langle x_{i}, - y_{N+1-j} \rangle =  \langle -x_{M+1-i}, y_{j} \rangle.
\end{equation*}
Besides, the input distribution $P_X(x_i) = \frac{1}{M} = P_X(x_{M+1-i})$, and the output distribution

\vspace{-.1in}
\begin{small}
\begin{equation*}
\begin{aligned}
    P_Y(y_j) 
    \!= \!& \int_{\mcX}\!\int_{\mcY_{N+1-j}}\! P_X(x) \frac{e^{-\frac{\| -y - Hx \|^2}{2\sigma_{Z}^{2}}}}{2\pi \sigma_{Z}^{2}}  \mrmd x\mrmd y 
    \!= P_Y(y_{N+1-j}),
\end{aligned}
\end{equation*}
\end{small}
\!\!\!\!
where $\mathcal{Y}_{j} = -\mathcal{Y}_{N+1-j}$ is the integral region to compute the probability mass function of the discrete point $y_i$.

In this way, we have proved that the marginal distribution is symmetric,
which allows us to further prove that
\begin{equation*}
\begin{aligned}
    \phi_{i}^{\ell} = & \phi_{M+1-i}^{\ell},~ 
    \psi_{j}^{\ell} =  \psi_{N+1-j}^{\ell},~ \forall i,j 
\end{aligned}
\end{equation*}
for arbitrary iteration number $\ell$ by the mathematical induction.

Firstly, we have
$\phi_{i}^{(0)} =1= \phi_{M+1-i}^{(0)}$ and $\psi_{j}^{(0)} =1= \psi_{N+1-j}^{(0)}$
due to the initial settings $\bm{\phi} = \mathbf{1}_{M}$ and $\bm{\psi} = \mathbf{1}_{N}$.

Next, assume $\phi_{i}^{\ell} = \phi_{M+1-i}^{\ell}$ and $\psi_{j}^{\ell} = \psi_{N+1-j}^{\ell}$ hold in the $\ell-$th iteration.
Then for the $(\ell+1)-$th iteration, we have

\vspace{-.1in}
\begin{small}
\begin{equation*}
\begin{aligned}
    \phi_{i}^{\ell + 1} 
    = \frac{P_X(x_{i})}  {\sum_{j = 1}^N \psi_{j}^{\ell} \mrme^{\lambda \langle x_{i}, y_{j} \rangle}}
    = \frac{P_X(x_{i})}  {\sum_{j = 1}^N \psi_{j}^{\ell} \mrme^{\lambda \langle x_{i}, y_{N+1-j} \rangle}}
    = \phi_{M+1-i}^{\ell + 1}.
\end{aligned}
\end{equation*}
\end{small}

Similarly, we can prove $\psi_{j}^{\ell + 1} = \psi_{N+1-j}^{\ell + 1}$.
Hence, the mathematical induction procedure is completed.

Moreover, we can conclude that
\begin{equation*}
\begin{aligned}
    & \sum_{i=1}^{M}\sum_{j=1}^{N} \langle x_i, y_j \rangle \phi_{i}^{\ell}\psi_{j}^{\ell} \\
    & = \frac{1}{2} \sum_{i=1}^{M}\sum_{j=1}^{N} \phi_{i}^{\ell}  \left( \psi_{j}^{\ell} \langle x_{i}, y_{j} \rangle + \psi_{N+1-j}^{\ell} \langle x_{i}, y_{N+1-j} \rangle \right)
    = 0.
\end{aligned}
\end{equation*}

As a result, when $H$ is positive define 
\footnote{In practice, the parameter $H$ in the AWGN channel is usually a combination of rotation and scaling transformation. And one could assume $H$ is positive define.}
, we have
\begin{equation*}
\begin{aligned}
    \hat{F}^{\ell}(0; \bm\phi, \bm\psi)
    = & \sum_{i=1}^M P_{X}(x_i) \langle Hx_i, x_i \rangle > 0,
\end{aligned}
\end{equation*}
which yields the inequality $F(0; \bm\phi, \bm\psi)>0$ always holds.
Noting that $F(\lambda; \bm\phi, \bm\psi) \rightarrow -T <0$ as $\lambda \rightarrow + \infty$,
the equation $F(\lambda; \bm\phi, \bm\psi)$ admits a unique solution on the interval $[0, \infty)$. 
\end{proofs}

\end{appendix}

\ifCLASSOPTIONcaptionsoff
  \newpage
\fi

\bibliographystyle{IEEEtran}

\bibliography{IEEEabrv}

\end{document}